\documentclass[useAMS,usenatbib]{mn2e}
\usepackage{amsmath,fleqn,graphicx,amssymb}
\usepackage{multirow}
\renewcommand{\[}{\begin{equation}}
\renewcommand{\]}{\end{equation}}
\def\p{\partial}

\def\ex#1{\left\langle#1\right\rangle}
\def\agama{{\sc agama}}
\def\github{{\sc github}}

{\newif\ifnotend
\notendtrue
\def\veclist{ABCDEFGHIJKLMNOPQRSTUVWXYZabcdefghijklmnopqrstuvwxyz.}
\def\top#1#2.{#1}
\def\tail#1#2.{#2.}
\loop\expandafter\xdef\csname v\expandafter\top\veclist\endcsname%
{{\noexpand\bf\expandafter\top\veclist}}
\edef\veclist{\expandafter\tail\veclist}
\if\veclist.\notendfalse\fi\ifnotend\repeat}
{\newif\ifnotend
\notendtrue
\def\veclist{ABCDEFGHIJKLMNOPQRSTUVWXYZ.}
\def\top#1#2.{#1}
\def\tail#1#2.{#2.}
\loop\expandafter\xdef\csname c\expandafter\top\veclist\endcsname%
{{\noexpand\cal\expandafter\top\veclist}}
\edef\veclist{\expandafter\tail\veclist}
\if\veclist.\notendfalse\fi\ifnotend\repeat}
\def\d{{\rm d}}

\def\Gyr{\,\mathrm{Gyr}}

\def\kpc{\,\mathrm{kpc}}

\def\kms{\,\mathrm{km\,s}^{-1}}
\def\km{\,\mathrm{km}}

\def\GeVdens{\,\mathrm{GeV\,cm}^{-3}}
\def\msun{\,{\rm M}_\odot}

\def\pc{\,\mathrm{pc}}
\def\e{\mathrm{e}}
\def\Rc{R_\mathrm{c}}\def\Rd{R_\mathrm{d}}
\def\fracj#1#2{{\textstyle{#1\over#2}}}

\def\jt{j_{\rm t}}
\title[Self-consistent models of our Galaxy]
{Self-consistent models of our Galaxy}

\author[James Binney \& Eugene Vasiliev]{
  James Binney$^1$\thanks{E-mail: binney@physics.ox.ac.uk} and Eugene
  Vasiliev$^{2}$\\  
  $^1$Rudolf Peierls Centre for Theoretical Physics, Clarendon Laboratory,
  Oxford, OX1 3PU, UK\\
  $^2$Institute of Astronomy, Madingley Road, Cambridge CB3 0HA
}

\begin{document}
\maketitle

\begin{abstract}
A new class of models of stellar discs is introduced and used to build a
self-consistent model of our Galaxy. The model is defined by the parameters
that specify the action-based distribution functions (DFs) $f(\vJ)$ of four stellar discs
(three thin-disc age cohorts and a thick disc), spheroidal bulge and
spheroidal stellar and dark haloes. From these DFs plus a specified
distribution of gas, we solve for the densities of stars and dark matter and
the potential they generate. The principal observational constraints are the
kinematics of stars with Gaia RVS data and the density of stars in the column
above the Sun. The model predicts the density and kinematics of stars and
dark matter throughout the Galaxy. We determine the structure of the dark
halo prior to the infall of baryons. A simple extension of the DFs of stellar
components to include chemistry allows the model to reproduce the way the
Galaxy's chemistry is observed to vary in the $(R,z)$ plane. Surprisingly,
the data indicate that high-$\alpha$ stars are confined to orbits with
$J_z\ga50\kpc\kms$.  The code used to create the model is available on
\github.
\end{abstract}

\def\Jd{J_{\rm d}} \def\Jv{J_{\rm v}}
\def\Jt{\widetilde J}
\def\Jdo{J_{\rm d0}} \def\Jvo{J_{\rm v0}}
\def\Jro{J_{\rm r0}} \def\Jzo{J_{\rm z0}} \def\Jpo{J_{\phi0}}
\section{Introduction} \label{sec:intro}

The volume and quality of the observational data that are available for our
Galaxy have increased spectacularly over the last decade. The
spectra of millions of stars have been taken from the ground, while ESA's
satellite Gaia has been tracking the motion across the sky of over a billion
stars, deriving for them photometry of unprecedented precision, and measuring
the line-of-sight velocities of millions of the brighter stars. The
astrophysical community now faces the challenge of synthesising these data
into a coherent  physical picture of how our archetypal  Galaxy is
structured, how it functions as a machine, and how it arrived at this state.

While ideally one would exploit the data for all the $>10^9$ stars that Gaia
has been tracking, many studies have focused on the $\sim7\times10^6$ stars
for which Gaia has measured line-of-sight velocities
\citep{Helmi2018,AntojaSpiral,HuntBubBovy2019,SellwoodTrick,TrickRix2019,AntojaDR3}.
We shall see that this subsample of the Gaia data is large enough for Poisson
noise to be insignificant, and it is relatively easy to model because when
the line-of-sight velocities are missing, one has to marginalise over them,
which is computationally expensive \citep[e.g.][]{LiBinneyRRLyrae}. The
potential strengths of the much larger sample of all Gaia stars are (i) that
it extends to much fainter magnitudes, and thus covers a more significant
slice of our Galaxy, and (ii) that it should be possible to determine its
selection function. Sadly, however, at the current time it is hard to
determine the probability that the data for a star of given magnitude, and
sky-coordinates will pass a given quality threshold
\citep{BoubertEverallb,BoubertEveralld}, and impossible to predict the
apparent magnitudes of distant stars of given absolute magnitudes on account
of the poorly known distribution of dust in the Galaxy
\citep[e.g.][]{LiBinneyYoungD}. These problems can be to a large extent
side-stepped by modelling the distribution of velocities at given locations
because the probability of a star entering Gaia's radial-velocity sample
(RVS) is independent of its velocity. In this paper we fit to Gaia RVS data
self-consistent, axisymmetric models of our Galaxy that are defined by
distribution functions (DFs) that are analytic functions of the action
integrals $J_i$.

Most previous fits of kinematic data have adopted a simple functional form for the
density of dark matter (DM) \citep{Roea03,JJB10,Bovy2013,Piea14}. This
procedure is unsatisfactory both because DM will actually respond in a non-trivial
way to the gravitational field of the stars, which is strongly flattened, and
because it yields no information regarding the velocity distribution of local
DM, which is important from the perspective of experiments to detect DM on
Earth. Here our approach is basically that pioneered by \cite{PifflPenoyreB} in
which a distribution function (DF) $f(\vJ)$ that depends on the action
integrals $J_i$ is assigned to both DM and various stellar populations,  and
then the gravitational field that stars and DM jointly generate is determined iteratively. 

Whereas \cite{PifflPenoyreB} simply computed one not very satisfactory model,
\cite{BinneyPiffl15} and \cite{ColeBinney} searched the space of candidate DFs
for ones that were consistent with data. The data they employed were rather
heterogeneous: astrometry of some stellar masers \citep{Reid2004}, terminal velocities of
neutral hydrogen and carbon monoxide \citep{Malhotra1995}, star counts from the Sloan Digital Sky
Survey \citep{SDSS1short} and the kinematics of giant stars measured  by the RAdial
Velocity Experiment \citep[RAVE;][]{RAVE1}. The heterogeneity of the data
combined with sub-optimal numerical methods made the search for an acceptable
model computationally expensive. 

Here we use the \agama\ software library \citep{AGAMA} to show that the data
for stars with Gaia line-of-sight velocities is so extensive that they
strongly constrain DFs for stars and DM on their own. We do this by
constructing fully self-consistent dynamical models of the Galaxy.  These
models are, for the first time, completely specified only by their DF: all
other quantities -- their density distributions, gravitational potential and
kinematics -- follow from the DF.  In Section \ref{sec:discDF} we describe a
new family of DFs for stellar discs, and in Section~\ref{sec:haloDF} we
describe the DFs we use to represent the bulge and the dark and stellar
haloes.  In Section \ref{sec:models} we compare one of our models with Gaia
kinematics and a variety of older data sources. This comparison confirms 
that the Galaxy's circular speed declines outwards at the solar radius and
yields values for the local densities of stars and dark matter. We infer the
structure of the dark halo before the baryons fell in under both the
assumption of adiabatic invariance and the assumption that the dark halo
originally had a central density cusp that was eliminated by baryons
upscattering dark-matter particles. In Section~\ref{sec:Hayden} we show that
reasonable hypotheses regarding the chemical compositions of the model's
stellar components provides a good fit to the distribution of stars in the
([$\alpha$/Fe], [Fe/H]) plane at various locations $(R,z)$ that were reported
by \cite{Hayden2015}.

\section{A new family of DFs for discs}\label{sec:discDF}

\cite{JJB12:dfs}, \cite{Bovy2013}, \cite{Piea14} and several others modelled the Galaxy's
discs with the quasi-isothermal DF that was introduced by \cite{JJB10} and modified to its
current form by \cite{JJBPJM11:dyn}. This DF was introduced in the context of
solar-neighbourhood kinematics using an analytic model of the Galactic potential $\Phi$.
Weaknesses in this DF emerged when \cite{PifflPenoyreB}, \cite{BinneyPiffl15} and
\cite{ColeBinney} used it while computing the self-consistently generated potential. They
employed ad-hoc work-arounds for these problems, but it is now time to address these
problems clinically and resolve them in a satisfying way. Our discussion
extends that in Section 4.4 of
\cite{AGAMA}.

The root problem with the quasi-isothermal DF is that it references the circular radius
$\Rc(J_\phi)$ and the radial and vertical epicycle frequencies $\kappa(J_\phi)$ and
$\nu(J_\phi)$. The way these quantities vary with $J_\phi$ depends on the potential, so
when they appear in the DF, the latter is no longer a function of the actions alone. This
dependence of the DF on $\Phi$ endangers the convergence of the algorithm \cite{JJB14}
introduced for finding the potential. The problem can be evaded by using the functions
$\Rc(J_\phi)$, etc., associated with a fixed but suitable potential instead of the real
potential, but this fix is inelegant and means that a model is not uniquely specified by a
set of DFs. 

Another problem with the quasi-isothermal DF is that it ceases to make sense physically
for orbits that are highly eccentric or strongly inclined to the plane. Therefore we now
write down an extension of the Exponential DF defined by \cite{AGAMA} that
does not reference $\Rd$, $\kappa$ or $\nu$ and is physically reasonable throughout
action space

Let $J_{\phi0}$, $\Jvo$ and $\Jdo$ be three fixed actions, the first much
larger than the other two. Then we define
\[\label{eq:defJvJd}
\Jv\equiv \vert J_\phi\vert +\Jvo\quad\hbox{and}\quad
\Jd\equiv \vert J_\phi\vert +\Jdo
\]
and take the DF of a disc component to be
\[\label{eq:trip_prod}
f(\vJ)=f_\phi(J_\phi)f_r(J_\phi,J_r)f_z(J_\phi,J_z),
\]
where the
functions $f_r$ and $f_z$ have the common form
\[\label{eq:defsJi}
f_i(J_\phi,J_i)=\left({\Jv\over J_{\phi0}}\right)^{p_i}\!\!\!
{1\over J_{i0}}\exp\left[-\left({\Jv\over J_{\phi0}}\right)^{p_i}\!\!\!{J_i\over
J_{i0}}\right] \quad (i={r,\,z}).
\]
Here the prefactor ensures that $1\simeq\int_0^\infty\d J_i\,f_i$. The factor
$(\Jv/J_{\phi0})^{p_i}$ in the exponential controls the radial gradient in the
velocity dispersion $\sigma_i$.

The function $f_\phi(J_\phi)$ in equation (\ref{eq:trip_prod}) determines the
radial structure of the disc. We adopt
\[\label{eq:fPhi}
f_\phi(J_\phi)\equiv
\begin{cases}
0&\hbox{when }J_\phi<0\cr
\displaystyle{{M\over(2\pi)^3}{J_\phi\over\Jpo^2}\e^{-\Jd/\Jpo}}&\hbox{if
}J_\phi>0.
\end{cases}
\]
The vanishing of $f$ for $J_\phi<0$ sets this DF apart from several earlier
DFs for discs, including that of \cite{AGAMA}, who proposed using
for $J_\phi<0$ the same formula as for $J_\phi>0$ multiplied by an additional
factor
\[
\exp\bigg({\Jv J_\phi\over J_{r0}}\bigg).
\]
When the DF at $J_\phi<0$ is taken to be a multiple of the DF at $J_\phi>0$
in this way,
one is implicitly assuming that there is a retrograde disc like the
prograde one but with reduced density. This is not the natural assumption:
rather we suppose that any disc stars that are now on retrograde orbits
have been scattered onto such orbits after being born on more-or-less
eccentric co-rotating orbits. Thus most retrograde stars should be on
highly inclined or eccentric orbits and an  implicit assumption that
the co-rotating disc is shadowed by 
a counter-rotating counterpart is unnatural. More natural is to deem any
counter-rotating stars to belong to a hot component, such as the bulge or the
stellar halo. Since the DF should be a continuous function on action space,
we arrange for the DF to go to zero as $J_\phi\to0+$ by replacing $\Jd$ in
the first line of
equation (21) in \cite{AGAMA} by $J_\phi$.

\begin{table}
\caption{Default  parameter values of the DF that a disc component. Actions are given in
$\!\kpc\kms$.}\label{tab:DF}
\begin{tabular}{ccccccc}
$J_{r0}$ & $J_{z0}$ & $J_{\phi0}$ & $p_r$ & $p_z$ &  $\Jvo$ & $\Jdo$\\
\hline
18 & 6 & 700 & 0.5 & 0.5 & 200 & 150\\
\end{tabular}
\end{table}

Here we present some examples of how the  characteristics of a disc
component vary with the parameters that specify its DF through equations
(\ref{eq:defJvJd}) to (\ref{eq:fPhi}). Table~\ref{tab:DF} lists the
default values taken by parameters that are not explicitly given in each example. 

\subsection{How the parameters affect the DF's values}

\begin{figure}
\centerline{\includegraphics[width=.9\hsize]{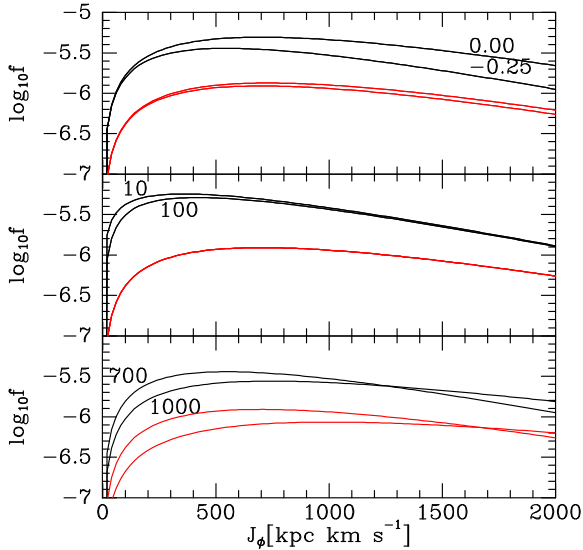}}
\caption{How the parameters $\Jpo$ (bottom panel),  $\Jv$ (middle panel) and
$p_i$ (top panel)
affect the DF. Black curves give values of the disc DF for $J_r=J_z=0$, while
red curves show values for generic actions ($J_r,J_z)=(10,5)\kpc\kms$). In
the bottom panel $\Jpo$ is set to $700$ and $1000\kpc\kms$ with the other
parameters as listed in Table~\ref{tab:DF}. In the middle panel  $\Jv$ is set
to $10$ and $100\kpc\kms$ with the other parameters as in
Table~\ref{tab:DF}. In the top panel $p_r=p_z$ are set to $-0.25$ and $0$.}\label{fig:JdJv}
\end{figure}

The scale actions $\Jro$ and $\Jzo$ set the scale on which the DF declines as
$J_r$ and $J_z$ increase: larger $\Jzo$ implies larger vertical velocity
dispersion $\sigma_z$, while increasing $\Jro$ increase both $\sigma_R$ and
$\sigma_\phi$.  In the bottom panel of Fig.~\ref{fig:JdJv} we plot in black
$f$ versus $J_\phi$ with $J_r=J_z=0$ for two values of $\Jpo$. Corresponding values for
$(J_r,J_z)=(10,5)\kpc\kms$ are plotted in red.
$\Jpo$ sets the scale over which the DF decreases as
$J_\phi$ grows. The steeper the decline, the smaller will be the scale length of
the resulting exponential disc, so $\Jpo$ is a surrogate for $\Rd$.

In the middle panel of Fig.~\ref{fig:JdJv}, the black and red curves are
again for $(J_r,J_z)=(0,0)$ and $(10,5)\kpc\kms$ but the curves in each
colour are now for $\Jvo=10$ and $100\kpc\kms$. We see that even a ten-fold
increase in $\Jvo$ has only a small effect on the value of the DF, so
$\Jvo$ is not an important
parameter: its role is to prevent the velocity dispersions diverging (for
$p_i>0$) or vanishing (if $p_i<0)$ at the centre.

The top panel of Fig.~\ref{fig:JdJv} shows how the DF varies with $J_\phi$
for two pairs of values, $(p_r,p_z)=(0,0)$ and $(-0.25,-0.25)$, of the
power-law exponents $p_r$ and $p_z$. Reducing the $p_i$ reduces the
difference between the slopes of the black curves for circular orbits
$(J_r=J_z=0)$ and the red curves for non-zero eccentricity and inclination. A
large separation between the red and black curves implies a DF that falls
rapidly with increasing eccentricity/inclination and thus small values of the
velocity dispersions $\sigma_R$ and $\sigma_z$.  Hence reducing the $p_i$
increases the velocity dispersions at large radii relative to their values at
small radii.

\subsection{Observational significance of parameters}

The real-space structure of a component depends on the
gravitational potential in which it resides. In general that potential
depends on all the model's components, so it will change with any component's
parameters. Here in the interests of clarity we fix the potential, choosing
the potential of a model that  provides a good fit to
observational data. Thus we adopt a very realistic potential and explore how
the structure of a single component depends on the parameters in its DF.

\begin{figure}
\centerline{\includegraphics[width=.8\hsize]{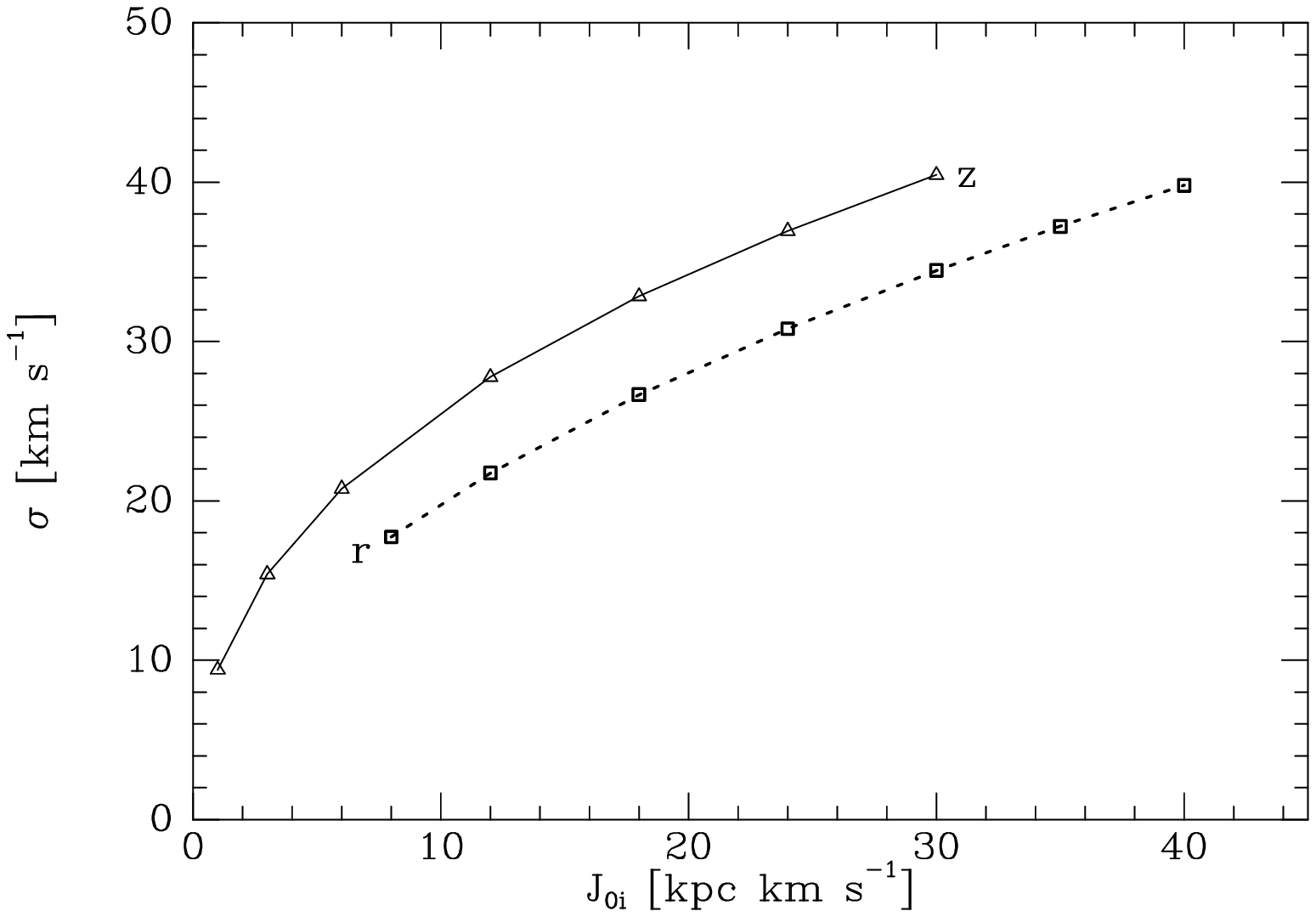}}
\centerline{\includegraphics[width=.8\hsize]{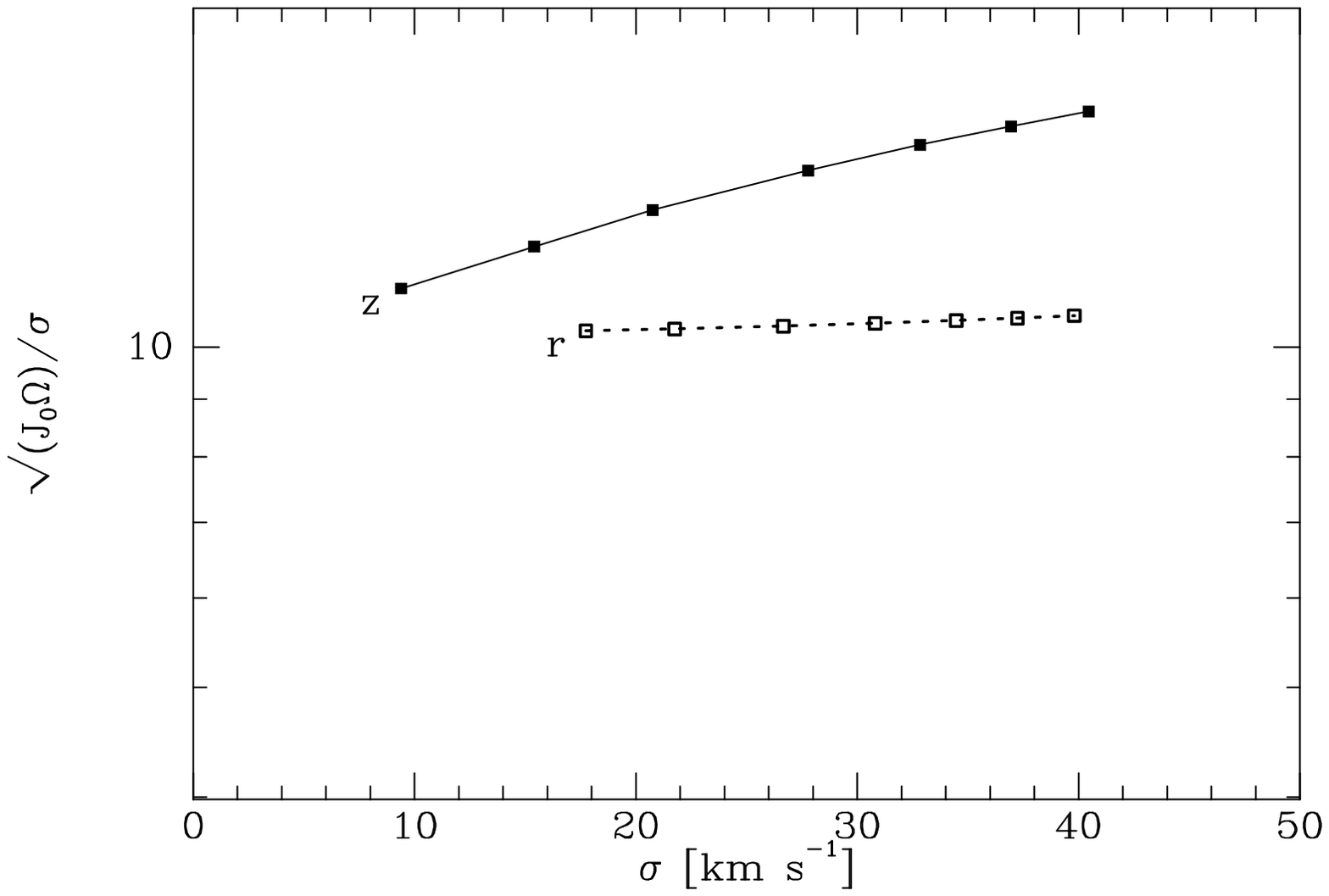}}
\caption{Upper panel: the radial (full line) and vertical (dashed line) velocity
dispersions at $R_0=8.27\kpc$ induced in a disc by the characteristic action
$J_{0i}$ ($i=r,z$) that appears
in equation (\ref{eq:defsJi}). Lower panel: the relationship between the velocity
dispersion (radial or vertical) in a disc and the square root of the product
of the characteristic action $J_{i0}$ in the DF and the mean orbital
frequency $\ex{\Omega_i}$.}\label{fig:sigzJz}
\end{figure}

\subsubsection{Roles of the characteristic actions $J_{\phi0}$, $J_{r0}$ and
$J_{z0}$}

The most important parameters are the three characteristic actions $J_{i0}$,
where $i=r,z,\phi$. While $J_{\phi0}$ sets the component's scale length,
$J_{r0}$ and $J_{z0}$ set the in-plane and vertical dispersions,
respectively. By setting these dispersions through $J_{r0}$ and $J_{z0}$, we
simultaneously determine a component's vertical density profile and
asymmetric drift through standard dynamics.

The upper panel of Fig.~\ref{fig:sigzJz} shows the relationship between $J_{r0}$ and
$J_{z0}$ and the velocity dispersions in the plane at $R_0=8.27\kpc$. The lower
panel shows that both dispersion satisfy quite accurately the relationship
\[\label{eq:omJ}
\sigma\simeq\sqrt{\ex{\Omega_i}J_{i0}},
\]
 where $\ex{\Omega_i}$ is the mean value of $\Omega_i$ of the given disc's
stars sampled at $(R,z)=(R_0,0)$. A relationship of this type is suggested by
the fact that $J_z\d\theta_z=v_z\d z+\d h$, where $h(\vx,\vv)$ is some
function on phase space: when we divide both sides by $\d t$ and average
along an orbit $\ex{\d h/\d t}=0$, so when we further average over orbits we
expect to find $J_z\ex{\Omega_z}=\ex{v_z^2}$. This argument indicates that in
the case $p_i=0$, when the exponentials in the DF make $\ex{J_i}$ the same
throughout action space, 
velocity dispersions should fall in parallel with frequencies. In particular
we expect $\sigma_z$ to fall with increasing distance from the plane, and
both $\sigma_R$ ad $\sigma_z$ to fall roughly as $1/R$ with distance from the
Galactic centre.

\begin{figure}
\centerline{\includegraphics[width=.8\hsize]{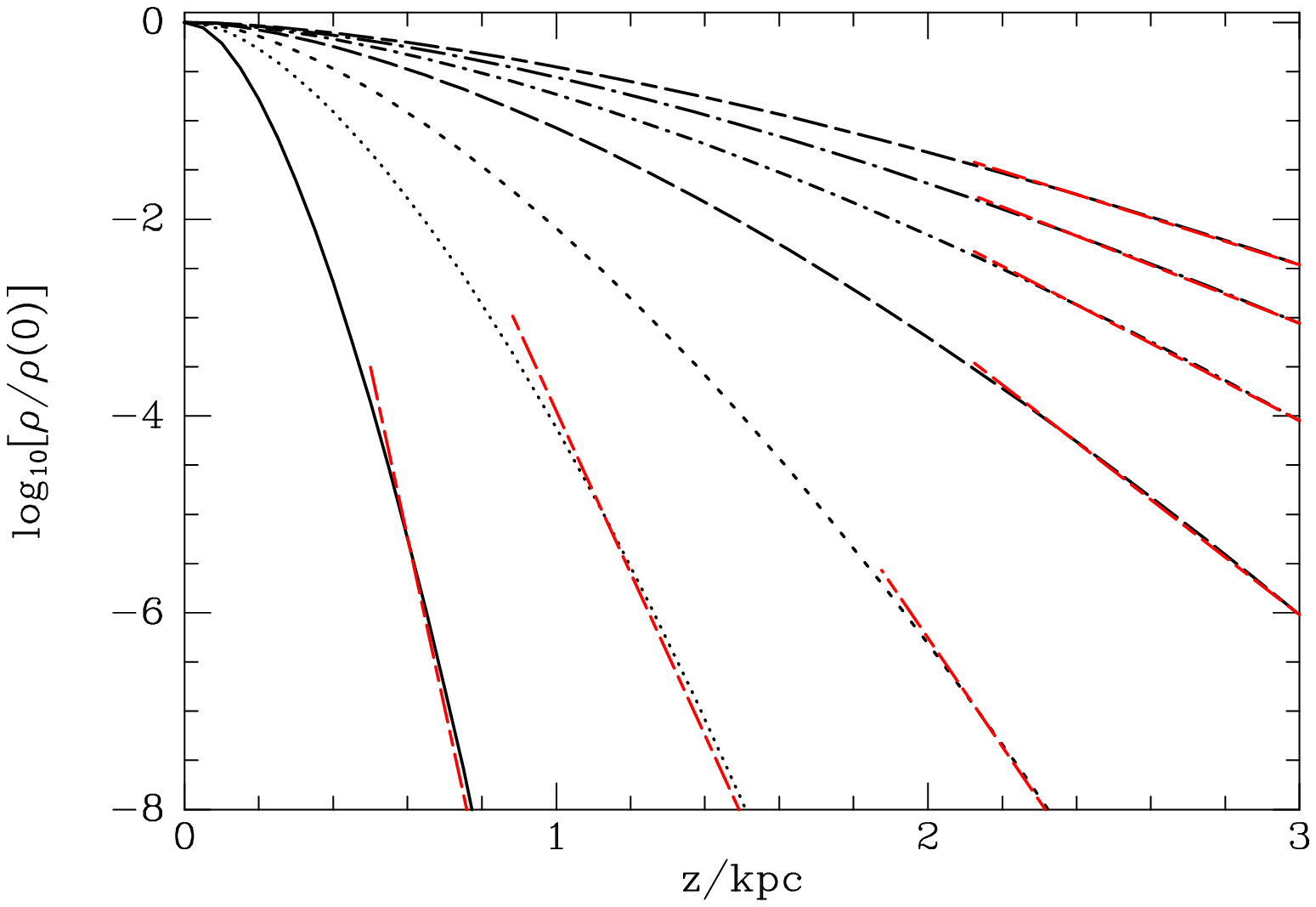}}
\centerline{\includegraphics[width=.8\hsize]{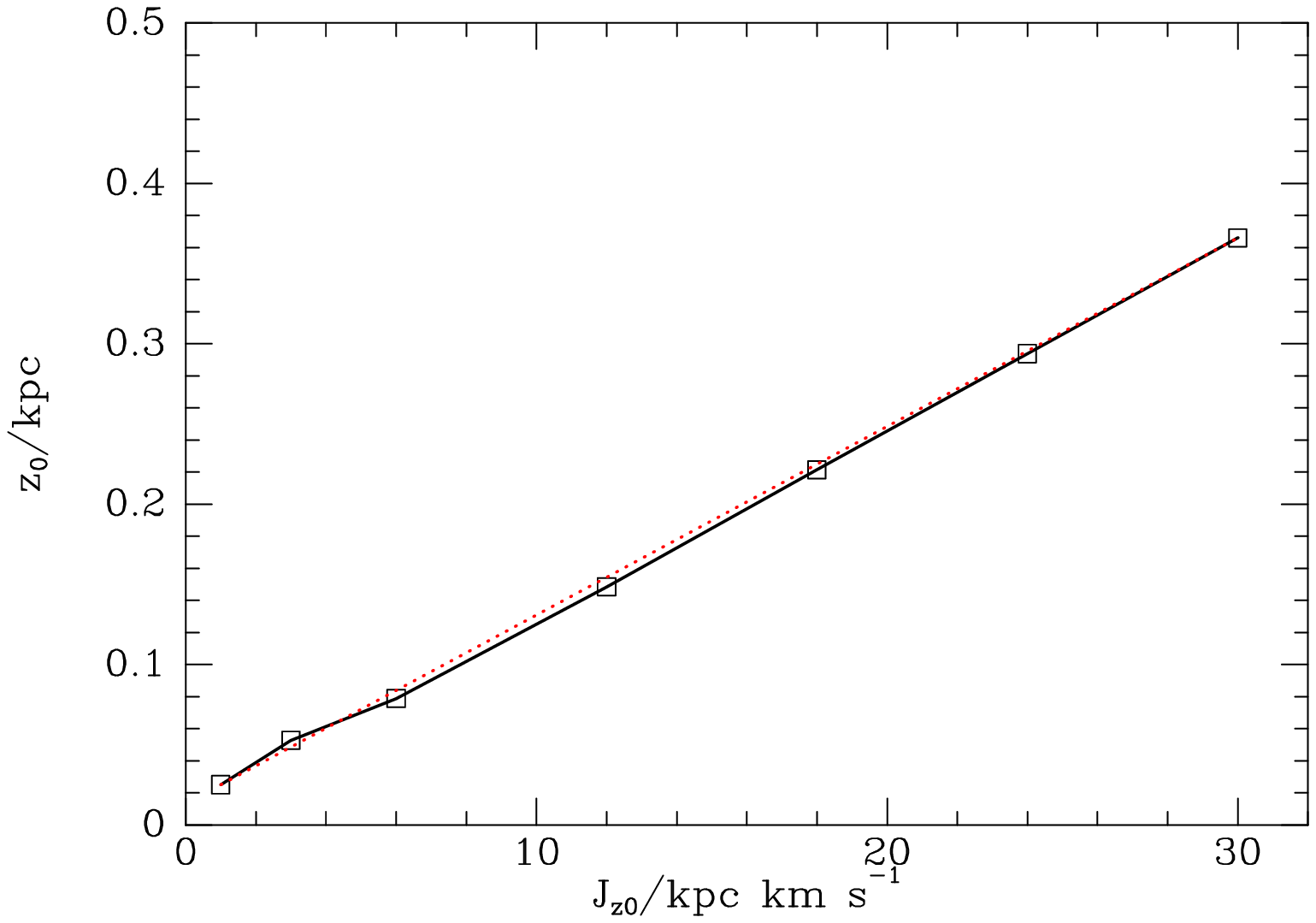}}
\caption{Upper panel: the vertical density profiles at $R_0=8.27\kpc$ in disc
components with $J_{z0}=1,3,6,12,18,24$ and $30\kpc\kms$. The red curves are
perfect exponentials fitted to the last eight points. Lower panel: asymptotic
scale height as a function of $J_{z0}$. The red dotted line has slope
$0.012\km^{-1}{\rm s}$}\label{fig:rhoZ}
\end{figure}

The upper panel of Fig.~\ref{fig:rhoZ} shows the vertical density profiles of the
components plotted in Fig.~\ref{fig:sigzJz}. The profiles steepen rapidly from zero in
the plane to a nearly perfect exponentials once the density has fallen to less than a tenth of
its central values.
The lower panel of Fig.~\ref{fig:rhoZ} plots the resulting asymptotic scale-height against
$J_{z0}$. The relationship is almost exactly linear:
\[
{z_0\over\!\kpc}\simeq0.012{J_{z0}\over\!\kpc\kms}.
\]

\begin{figure}
\centerline{\includegraphics[width=.8\hsize]{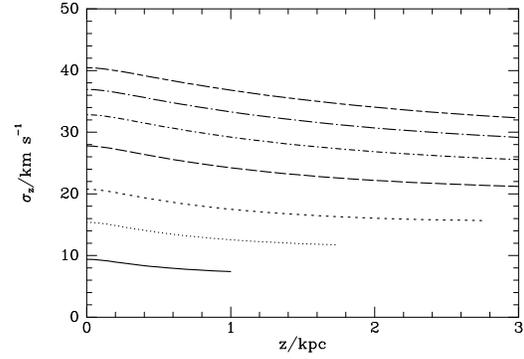}}
\caption{The variation at $R_0=8.27\kpc$ with $z$ of the vertical velocity
dispersion of disc components with $J_{z0}=1,3,6,12,18,24$ and $30\kpc\kms$.
The curves are terminated when $\rho(z)/\rho(0)$ falls to
$10^{-10}$.}\label{fig:sigZ}
\end{figure}

\begin{figure}
\centerline{\includegraphics[width=.8\hsize]{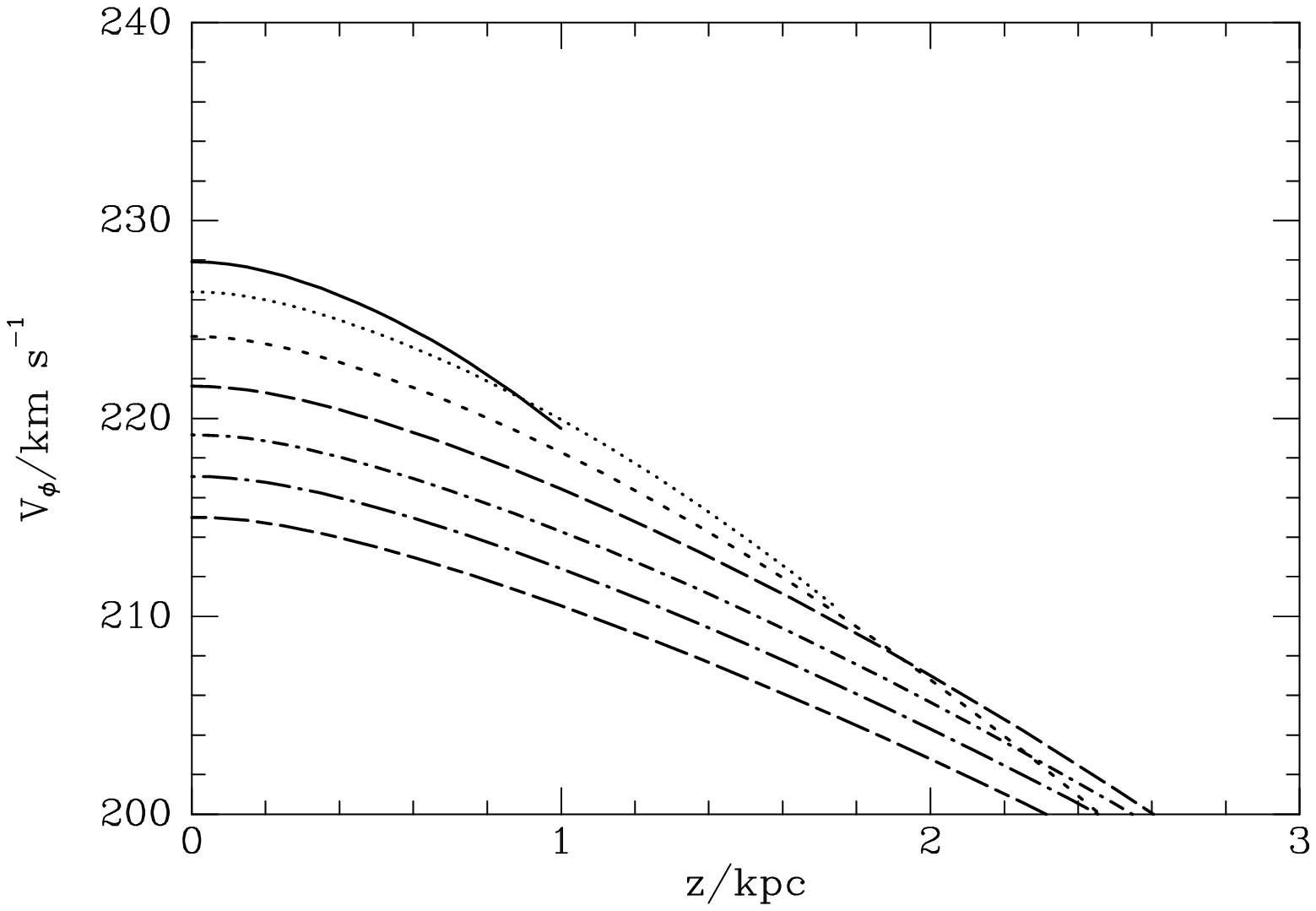}}
\centerline{\includegraphics[width=.8\hsize]{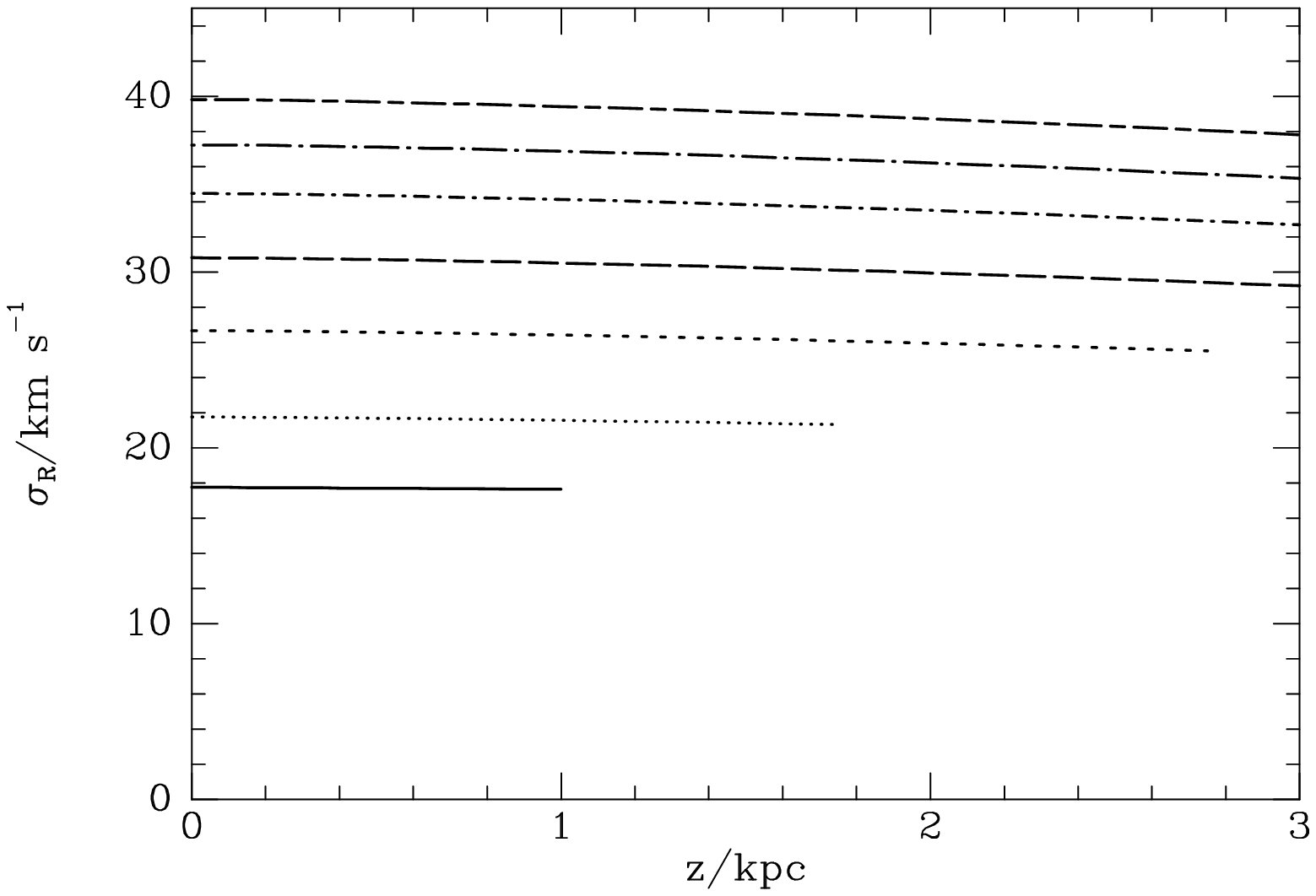}}
\caption{The variation at $R_0=8.27\kpc$ with $z$ of $\ex{V_\phi}$ (upper panel) and
$\sigma_R$ (lower panel) in increasingly hot disc 
components: from top to bottom in  the upper panel the characteristic actions
are $(J_{r0}, J_{z0})=(8,1),(12,3),(18,6),(24,12),(30,18),(35,24)$
and $(40,30)\kpc\kms$. The  curves are
terminated when $\rho(z)/\rho(0)$ falls to $10^{-10}$.}\label{fig:sigVSZ}
\end{figure}

Fig.~\ref{fig:sigZ} shows how the vertical velocity dispersion in a disc
component varies with height above the plane.  The dispersion falls slowly as
expected from the argument above relating $\sigma_z$ to the mean frequency
$\Omega_z$, while the density declines by orders of magnitude.
The decline in $\sigma_z$ is steepest near the mid-plane.

Fig.~\ref{fig:sigVSZ} shows the mean rotation speed (upper panel) and the radial velocity
dispersion (lower panel) in a disc component at $R_0=8.27\kpc$ as functions of height. The
radial velocity dispersions are to high precision independent of height. The thinnest
components have, by construction, the smallest dispersions, so at $z=0$ they rotate fastest
-- they show the least asymmetric drift. In every component $\ex{V_\phi}$ falls
appreciably with height.

\begin{figure}
\centerline{\includegraphics[width=.8\hsize]{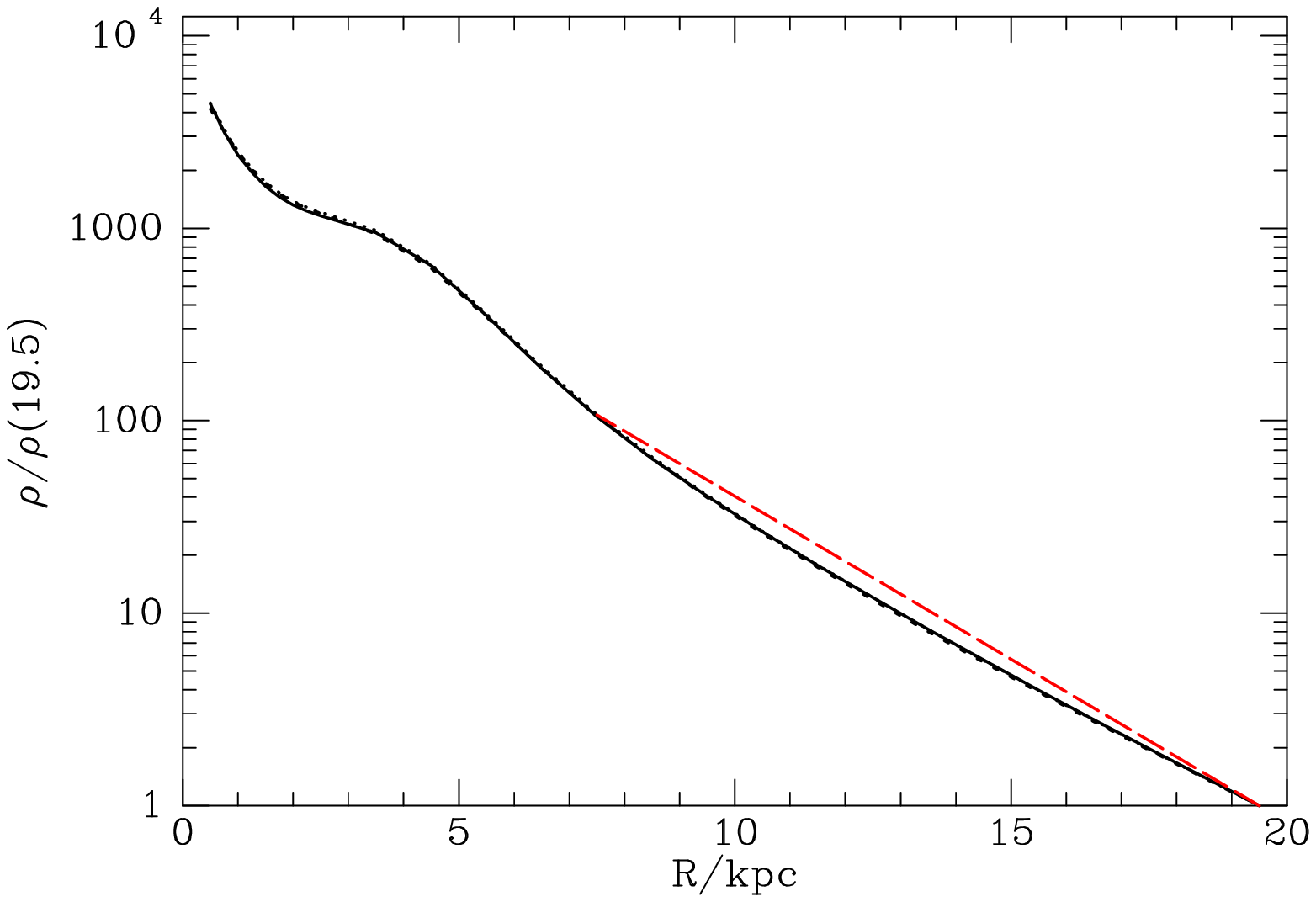}}
\centerline{\includegraphics[width=.8\hsize]{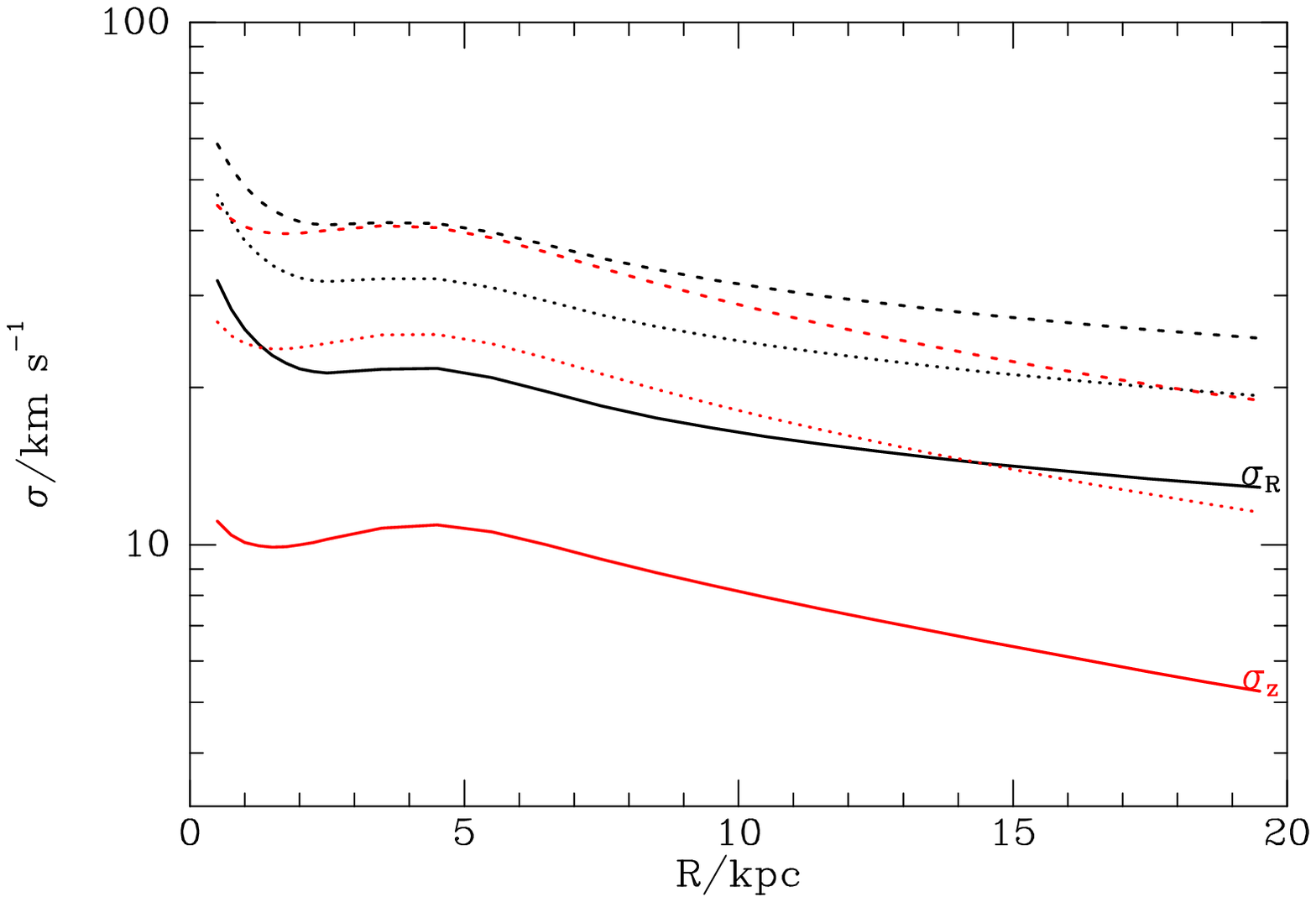}}
\caption{Upper panel: the variation with $R$ of the density at $z=0$ of disc
components with $(J_{r0},J_{z0})=(8,1),(18,6)$ and $(30,18)\kpc\kms$. The
dashed red line shows an exponential decline with scale-length $2.72\kpc$.
Lower panel: the mid-plane velocity dispersions of these components in the
radial (black) and vertical (red) directions.}\label{fig:SDR}
\end{figure}

The upper panel of Fig.~\ref{fig:SDR} shows the radial density profiles of
the disc components with $(J_{r0},J_{z0})=(8,1),(18,6)$ and
$(30,18)\kpc\kms$. Beyond $R_0$ these profiles are quite accurately
exponential with the common scale length $R_{\rm d}=2.72\kpc$ that is
indicated by the red dashed line. This scale-length is set by the value
$J_{\phi0}=700\kpc\kms$ assigned to all components. Interior to $R_0$ the
profiles acquire structure from the bulge via the self-consistently generated
potential, as will be described below.  Despite this structure, the profiles
approximate exponentials at all $R$.

The lower panel of Fig.~\ref{fig:SDR} shows the dependence on radius of the
velocity dispersions of these components in the radial (black) and vertical
(red) directions. The ratio $\sigma_z/\sigma_R$ increases rapidly with disc
thickness, approaching unity for the hottest disc plotted [which has
$(J_{r0},J_{z0})=(35,24)\kpc\kms$].  This result is a natural consequence of the
decision to have $J_{r0}$ span a smaller range ($8-40\kpc\kms$) than
$J_{z0}$ ($1-30\kpc\kms$). In the thinnest disc (full curves) the vertical
dispersion falls almost exponentially with $R$, while the radial dispersion
falls more slowly. This result reflects the fact that $\Omega_z$ is much more
sensitive than $\Omega_r$ to the contribution of the disc to the overall
gravitational field: around $R_0$ the density of the disc that generates the
gravitational field used for these figures is declining nearly exponentially
with $R$.

\begin{figure}
\centerline{\includegraphics[width=.8\hsize]{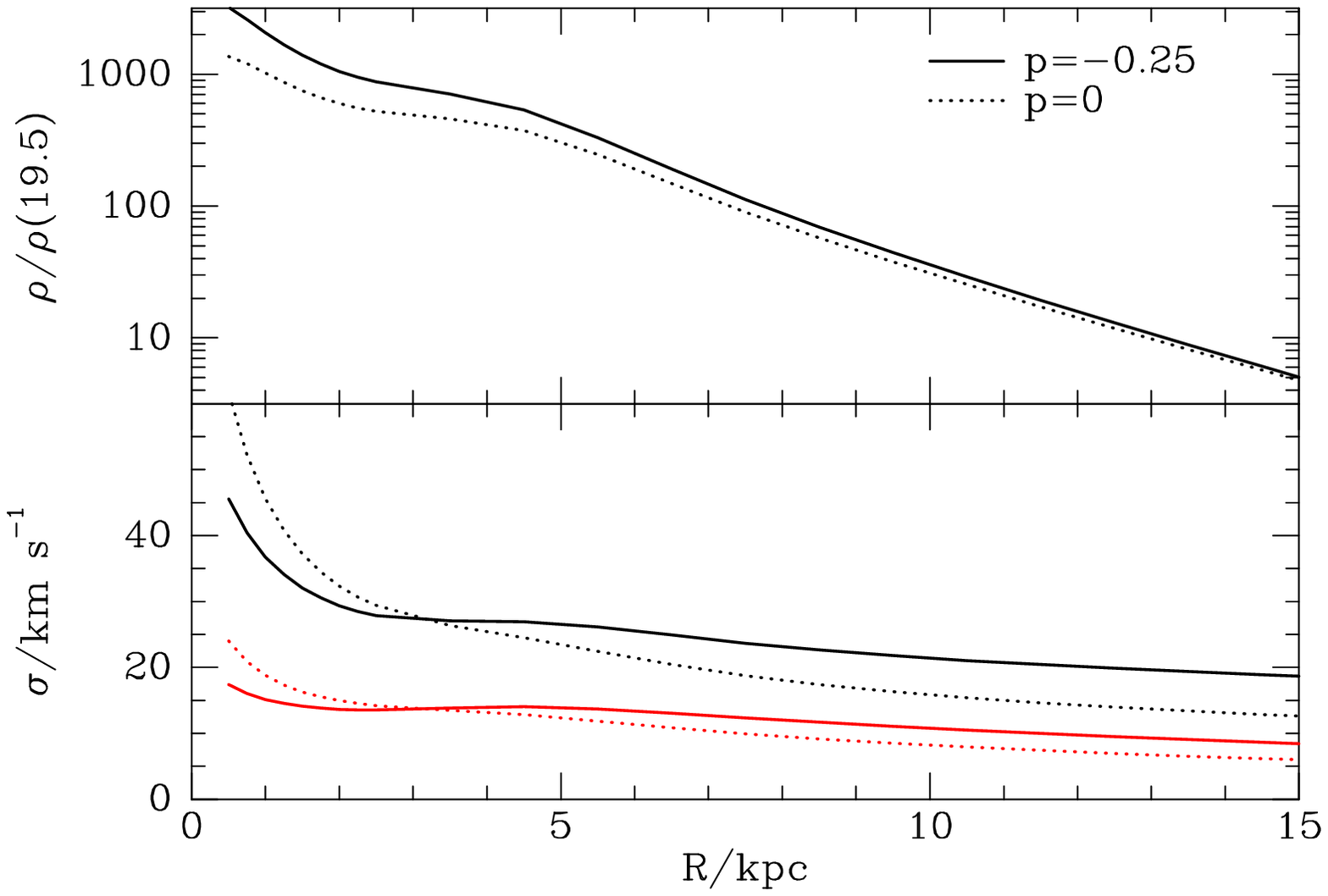}}
\centerline{\includegraphics[width=.8\hsize]{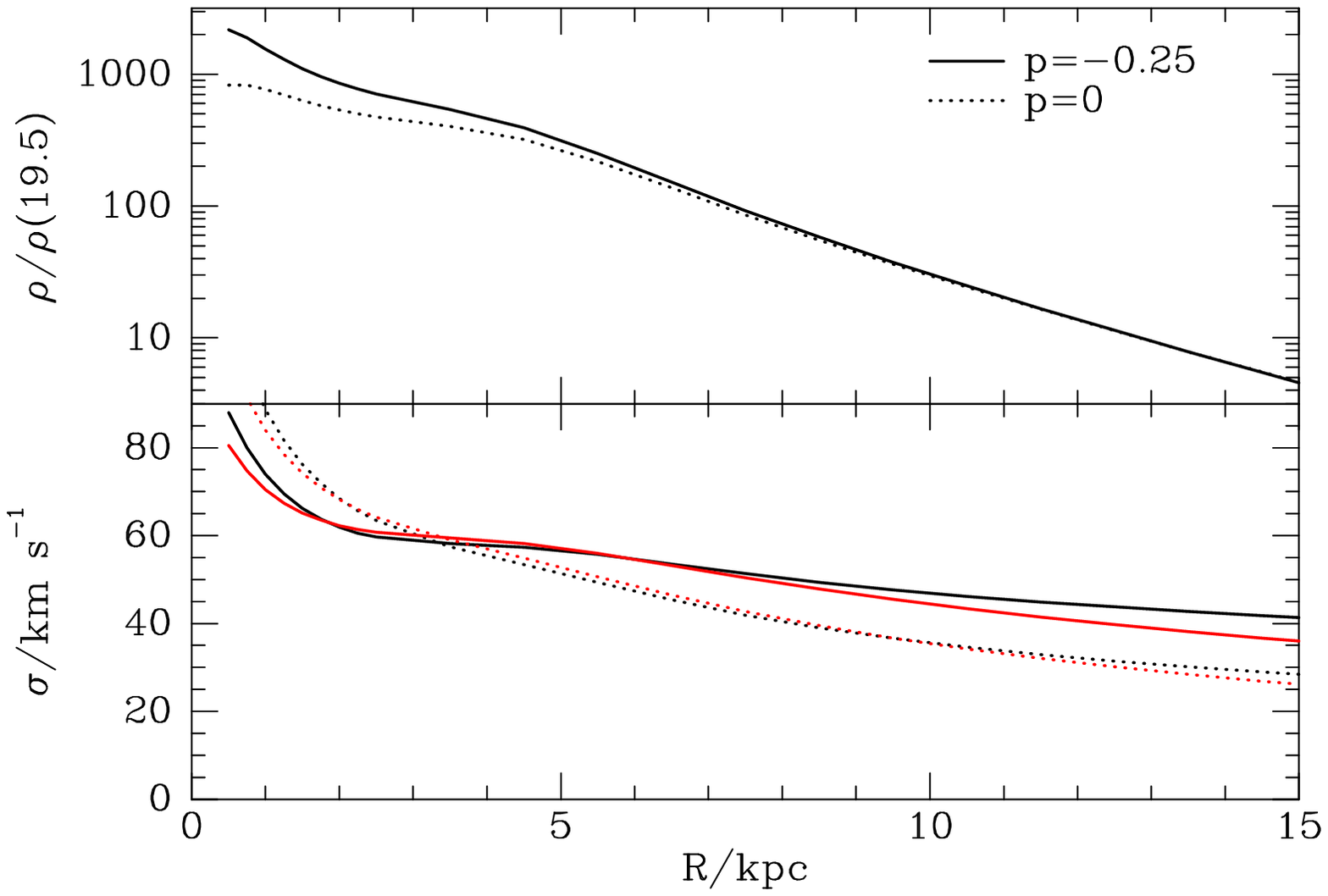}}
\caption{The real-space consequences  of changing the exponents $p_i$ in 
discs that are cold (top) and hot (bottom). The cold disc has
$(J_{r0},J_{z0})=(8,1)\kpc\kms$, while the hot disc has
$(J_{r0},J_{z0})=(30,40)\kpc\kms$ and in all cases $p_r=p_z$. Full curves are
for $p_i=-0.25$, which yields a slower decline in velocity dispersions with
$R$ than the larger value of $p_i$.}\label{fig:powers}
\end{figure}

\subsubsection{Role of the exponents $p_r$ and $p_z$}

The exponents $p_r$ and $p_z$ that occur in equation (\ref{eq:defsJi}) moderate
the rates at which the $\sigma_R$ and $\sigma_z$ decrease outwards. The
results plotted above are for $p_i=0$. We will find that data for our Galaxy
requires slightly negative values of $p_i$, which cause $\sigma_R$ and
$\sigma_z$ to decrease outwards more slowly.  

In Fig.~\ref{fig:powers} we
show the mid-plane density and velocity dispersions for $p_i=-0.25$ (full
curves) and $p_i=0$ in the case of a cold disc and a hot disc (lower pair of
panels). Larger values of $p_i$ correspond to steeper outward gradients of
the dispersions, as expected.

\begin{figure}
\centerline{\includegraphics[width=.8\hsize]{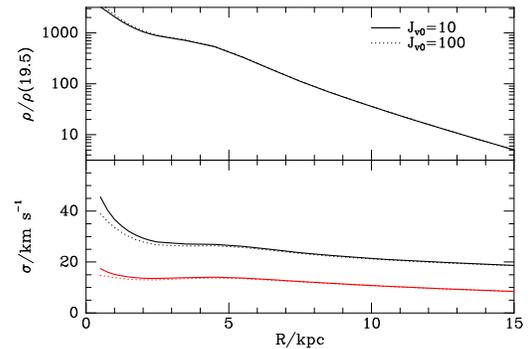}}
\caption{Real-space consequences of increasing the parameter $\Jvo$. Full
curves are for $\Jvo=10\kpc\kms$ and the barely visible dotted curves are for
$\Jvo=100\kpc\kms$. These
results are for a cold disc $(J_{r0},J_{z0})=(8,1)\kpc\kms$; increasing
$\Jvo$ has an even smaller impact on a hot disc.}\label{fig:Jv}
\end{figure}

\subsubsection{Role of the characteristic action $\Jvo$}

Fig.~\ref{fig:Jv} shows the effect on a cold disc of increasing the
characteristic action $\Jvo$ from $10$ to $100\kpc\kms$. This parameter is
active only for non-zero $p_i$, so Fig.~\ref{fig:Jv} compares profiles for a
disc with $p_i=-0.5$. As expected, $\Jvo$ has a perceptible effect only near
the centre and in the sense that larger values lower the central velocity
dispersions, and by hydrostatic balance, correspondingly increase the central
mid-plane density. However, the effect on the plotted
cold disc of even ten-fold decrease in $\Jvo$ is extremely small. The effect
is no larger in the case of a hot disc.

\section{Halo DFs}\label{sec:haloDF}

To build a self-consistent model Galaxy we require DFs for at least two hot
components: the dark halo and the bulge / stellar halo. On account of the
Galactic bar, a satisfactory model
of the bulge would be non-axisymmetric so cannot be provided with our
technology and we are obliged to model the bulge as an axisymmetric
structure. Since few bulge stars reach the volume around the Sun in which the
Gaia data are most precise, and the non-axisymmetric components of the
bulge's gravitational field decay strongly with increasing Galactocentric
radius, we anticipate obtaining useful results even with a completely
axisymmetric model.

For all three hot components of our Galaxy we employ a modified double
power-law DF that is related to, but different from, those introduced by
\cite{Poea15} -- hereafter P15. Our modifications of the P15 proposal are two-fold.
First the phase-space density at the origin of action space is made finite by
the mechanism introduced by \cite{ColeBinney}. Second, the linear combination
of actions $h=J_r+\eta_z J_z+\eta_\phi \vert J_\phi\vert $ that forms the
basis of the P15 proposal is replaced by a more complex
function of the actions. A full explanation of this replacement lies beyond
the scope of this paper and will be presented elsewhere
\citep{Binney_in_prep} but it addresses a weakness of the P15 proposal that was
mentioned already by \cite{BinneyPiffl15}, namely unphysical
behaviour of velocity distributions at small $\vert V_\phi\vert $.

The P15 halo DF  is
\[\label{eq:dpowerf}
f={M\over(2\pi J_0)^3}
{(1+J_0/h)^{\alpha}\over(1+g/J_0)^{\beta}}\exp(-[g/J_{\rm
cut}]^\delta).
\]
Instead of  the linear combinations $g(\vJ)$ and $h(\vJ)$, we use
\[\label{eq:defg}
g(\vJ)=h(\vJ)=J_r\e^{-\beta\sin(c\pi/2)}+\fracj12(1+c\xi)\e^{\beta\sin(c\pi/2)}\cL.
\]
Here
\[
c\equiv{L\over L+J_r}
\]
with $L\equiv J_z+\vert J_\phi\vert $ is a measure of an orbit's circularity and
\[
\cL(J_z,J_\phi)\equiv aL_z+b{J_z\vert J_\phi\vert \over L}+\vert J_\phi\vert 
\]
is a generalisation of the total angular momentum.  The constants $a$ and $b$
in the definition of $\cL$ are chosen such that $\p\cL/\p V_\phi$ vanishes as
$V_\phi\to0$. Specifically
\[
a=\fracj12(k+1)\hbox{ and }b=\fracj12(k-1),
\]
where $k$ is a number grater or equal to unity. We allow $k$ to depend on
energy by writing 
\[\label{eq:defk}
k(\xi)=(1-\xi)F_{\rm in}+\xi F_{\rm out}
\]
where $F_{\rm in}$ and $F_{\rm out}$ are constants and  $\xi$ is a
dimensionless surrogate for energy that increases from zero at the origin of
phase space to unity for marginally bound orbits through the formula
\[
\xi\equiv {\jt^\alpha\over\jt^\alpha+1}
\]
where $\alpha\simeq0.6$ and
\[
\jt\equiv{1.5J_r+L\over L_0},
\]
where $L_0\simeq6J_0$ is a scale action.

When $F_{\rm in}=F_{\rm out}=1$, $\cL=L$ and in a spherical potential the
distributions of $V_\theta$ and $V_\phi$ are identical. If 
$\alpha$ and $L_0$ are then set to the values suggested above, the velocity
distribution becomes isotropic when $\beta=0$ and radially biased when
$\beta>0$. To flatten the component in a spherical potential, one sets $F_{\rm in}$ and $F_{\rm
out}$ above unity. If the potential is flattened, setting $\beta=0$, and $F_{\rm
in}$ and $F_{\rm out}$ to the ratio $\nu/\Omega$ of vertical to azimuthal
epicycle frequencies at small and large radii, respectively causes the
velocity distribution to be nearly isotropic.

\section{Models of our Galaxy}\label{sec:models}

In this section  we use the DFs for disc components defined above to fit multi-component,
self-consistent models to data for stars in the Gaia DR2 release
\citep{GaiaDR2general} for which the Radial Velocity Spectrometer (RVS)
measured line-of-sight velocities \citep{GaiaKatz2018}.

\subsection{The data} 

We used the stellar locations and velocities computed by
\cite{SchoenrichME2019} for the RVS stars. These authors adopted $R_0=8.27\kpc$
as the distance to the Galactic centre and
$(U_0,V_0,W_0)=(11.1,250,7.47)\kms$ as the Sun's Galactocentric velocity.
According to this assumption, the Sun is moving towards the Galactic centre,
ahead of the local circular speed $\Theta_0=239\kms$ and up out of the plane.
They determined a distance-dependent selection function of the RVS sample and
used the kinematic method of \cite{SchoenrichBA} to determine the offset of
Gaia DR2 parallaxes. After determining this offset, they obtained moments for
the probability distribution of the distance to each star using the Bayesian
technique of \cite{SchoenrichAumer2017}.

We restricted the sample to stars with quoted uncertainties
$\sigma_\varpi<\varpi/5$ smaller than a fifth of their parallax $\varpi$.
The selection function of the RVS sample is complex \citep{BoubertEveralld} and a
decision was taken not to engage with it in this preliminary work. From this
decision it follows that we cannot attach significance to the {\it number} of
stars listed in any spatial bin; only the distribution of these stars in
velocity is significant. 

\begin{table}
\caption{The boundaries (kpc) of the bins in the $(R,\vert z\vert )$ plane used to
construct velocity histograms from Gaia data and models.} \label{tab:bdys}
\begin{tabular}{ccccccccc}
\hline
$R-R_0$&$-3$&$-1.5$&$-0.5$&$0.5$&$1.5$&$3$\\
$|z|$&0&0.1&0.2&0.35&0.5&1&1.75&3\\
\hline
\end{tabular}
\end{table}

We used the data given by \cite{SchoenrichME2019} to compute the histograms
of Galactocentric velocity components $V_R$, $V_z$ and $V_\phi$ for each of
35 spatial bins. These bins are defined by a rectangular grid in the
$(R,|z|)$ plane that stretches from $R_0-3\kpc$ to $R_0+3\kpc$ and from the
plane up to $|z|=3\kpc$.  Table~\ref{tab:bdys} defines this grid.

\begin{figure*}
\includegraphics[width=.8\hsize]{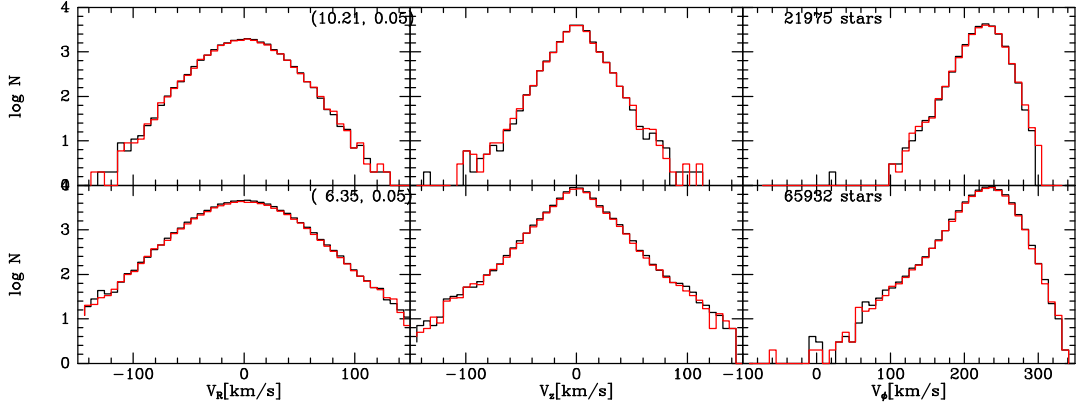}
\caption{The impact of observational uncertainty on velocity histograms. The
black histograms show the result of plotting a sample drawn from a realistic
model without allowing for errors, while the red histograms are obtained
from the same sample after scattering by typical Gaia DR2
errors. Both rows are for bins with $|z|<0.1\kpc$ but  the upper row is
for a bin at $R-R_0\sim2\kpc$, while the lower row is for the bin at
$R-R_0\sim-2\kpc$.}\label{fig:scatter}
\end{figure*}

\subsection{Observational uncertainties}

We investigated the impact of observational uncertainties on the histograms
by drawing a sample from a realistic model using the approximation to the RVS
selection function given by \cite{SchoenrichME2019} and comparing the
histograms one obtains from the raw sample and from the sample after the
stars have been scattered by the uncertainties in distance, proper motion and
line-of-sight velocity. The uncertainties in each observable were averages of
the DR2 uncertainties given for stars that lay at essentially the same
distance from us -- we used 50 equal-width bins in distance out to
$3\kpc$. We checked the correctness of the code  by computing the
differences between the observables before and after scattering and showing
that the standard deviations of these differences agree with the
uncertainties used to scatter the stars.

Errors in velocities will be most important for the most distant stars, so
Fig.~\ref{fig:scatter} shows their effect on the most distant bins, looking
inward in the lower row of panels and outward in the upper row.  The red
histograms computed after scattering stars essentially obliterate the black
histograms computed before scattering the raw sample.  This perhaps
surprising result allows us to neglect observational uncertainties even when
using DR2 data, and in the following we compare histograms from Gaia with the
model's distribution of velocities at the barycentre of the Gaia stars in
each bin. EDR3 data is superior to DR2 data in that its astrometric
uncertainties are smaller, so if we upgraded to EDR3 data, more stars would
pass our quality cuts so more stars would contribute to each histogram, and
observational uncertainties would be even less significant. But given that
our data/model comparisons are limited neither by Poisson noise nor by
observational uncertainties, we continue to use the DR2 data. 

\subsection{The models}

Our Galaxy models comprise eight components of which seven are defined by DFs
and the eighth, the gas disc, is defined as a  density distribution
that contributes to the total potential. Following \cite{WDJJB98:Mass} the
gas disc has  density
\[\label{eq:gas_d}
\rho_{\rm g}(R)={\Sigma_0\over 4z_0}\e^{-R/R_\d-R_{\rm
h}/R}\hbox{sech}^2\Big({z\over2z_0}\Big).
\]
with $\Sigma_0=1.3\times10^8\msun\kpc^{-2}$, $R_\d=5\kpc$, $R_{\rm h}=5\kpc$
and $z_0=60\pc$. This gas disc has total mass $1.04\times10^{10}\msun$, and
at the solar radius $R_0=8.27\kpc$ has surface density
$\Sigma_g(R_0)=13.6\msun\pc^{-2}$ and central density
$\rho_g(R_0,0)=0.057\msun\pc^{-3}$.

We include three spheroidal
components -- a dark halo, a bulge and a stellar halo -- and four disc
components -- a thick disc and old ($\tau>6\Gyr$), middle-aged
($6>\tau>2\Gyr$), and young  components ($\tau<2\Gyr$) of the
thin disc. Each of these discs has a DF of the type defined in
Section~\ref{sec:discDF}, while the spheroidal components have DFs of the type
defined in Section~\ref{sec:haloDF}.

\begin{table*}
\caption{DFs fitted to Gaia data. Normalisations in $10^{10}\msun$, 
actions in $\kpc\kms$}\label{tab:DF_G}
\begin{tabular}{lcccccccccc}
Spheroids&Mass&$J_c$&$J_0$&$J_{\rm cut}$&$\alpha$&$\beta$&$\delta$&
$F_{\rm in}$&$F_{\rm out}$&$L_0$\\
\hline
Dark halo    &98&100&10000&20000&1.6&2.7&2&1.4&1.2&100000\\
Stellar halo&0.1&  5&600  &100000&1 &3.5&2&1.8&1.2&100000\\
Bulge       &1.1&  5& 19.5&200  &0.5&1.8&2&  3&  2&100000\\
\hline
\\
Discs&Mass&$J_{\phi0}$&$J_{r0}$&$J_{z0}$&$p_r$&$p_z$&$\Jdo$&$\Jvo$\\
\hline
Young disc        &0.175&650&10&0.65&-0.7&-0.3&10&100\\
Middle disc       &0.75&600&17&2.8&-0.35&-0.1 &10&700\\
Old disc          &1.0&550&22&5&-0.25&-0.1 &10&700\\
Thick disc        &0.95&400&63&30&0.13&0.05& 20& 40\\
\end{tabular}
\end{table*}

\begin{figure*}
\centerline{\includegraphics[width=.8\hsize]{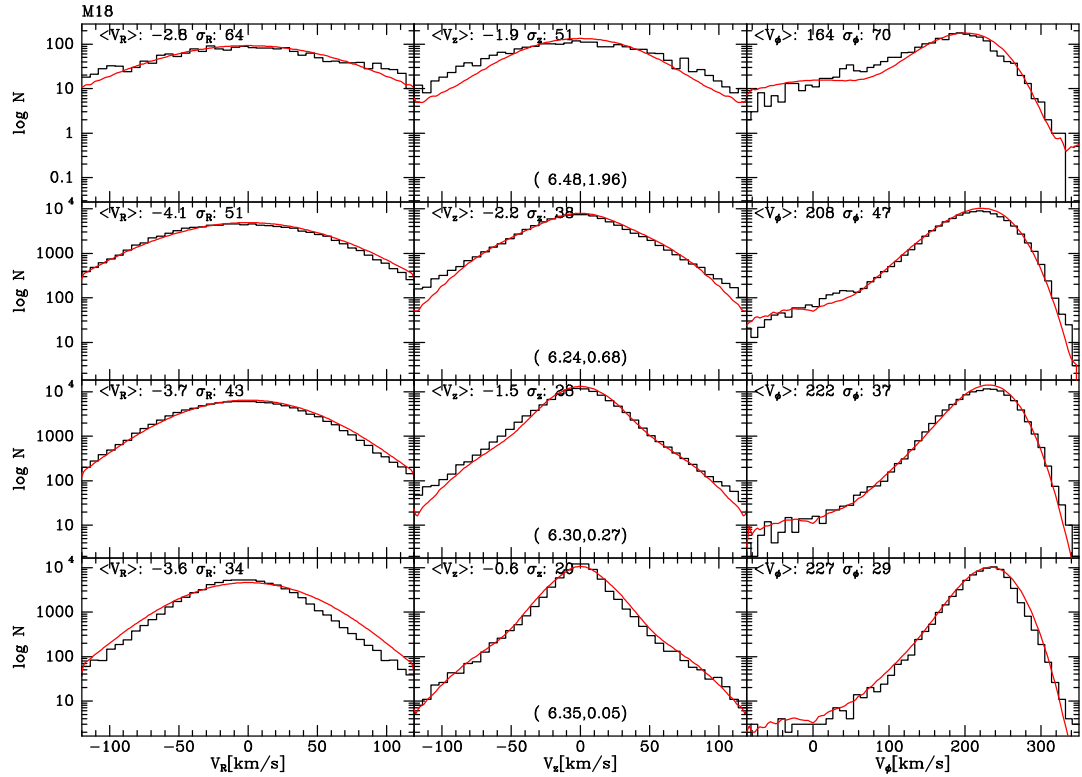}}
\caption{Black histograms show Gaia data for stars with radii $R-R_0\in(-3,-1.5)\kpc$ and 
distances from the plane that increase from bottom to top (barycentres of the cells are
given at upper right of the cells in the first two columns). Means and standard deviations of
the velocities are given at top left of each panel. From left to right the columns are for
$V_R$, $V_z$ and $V_\phi$. The red curves show the predictions of the dynamical model with
the total number of stars in a cell normalised to agree with the data. No allowance has been made
for observational errors. M18 at top left in this and subsequent figures is
simply the model's designation.}\label{fig:Rin}
\end{figure*}

\begin{figure*}
\centerline{\includegraphics[width=.8\hsize]{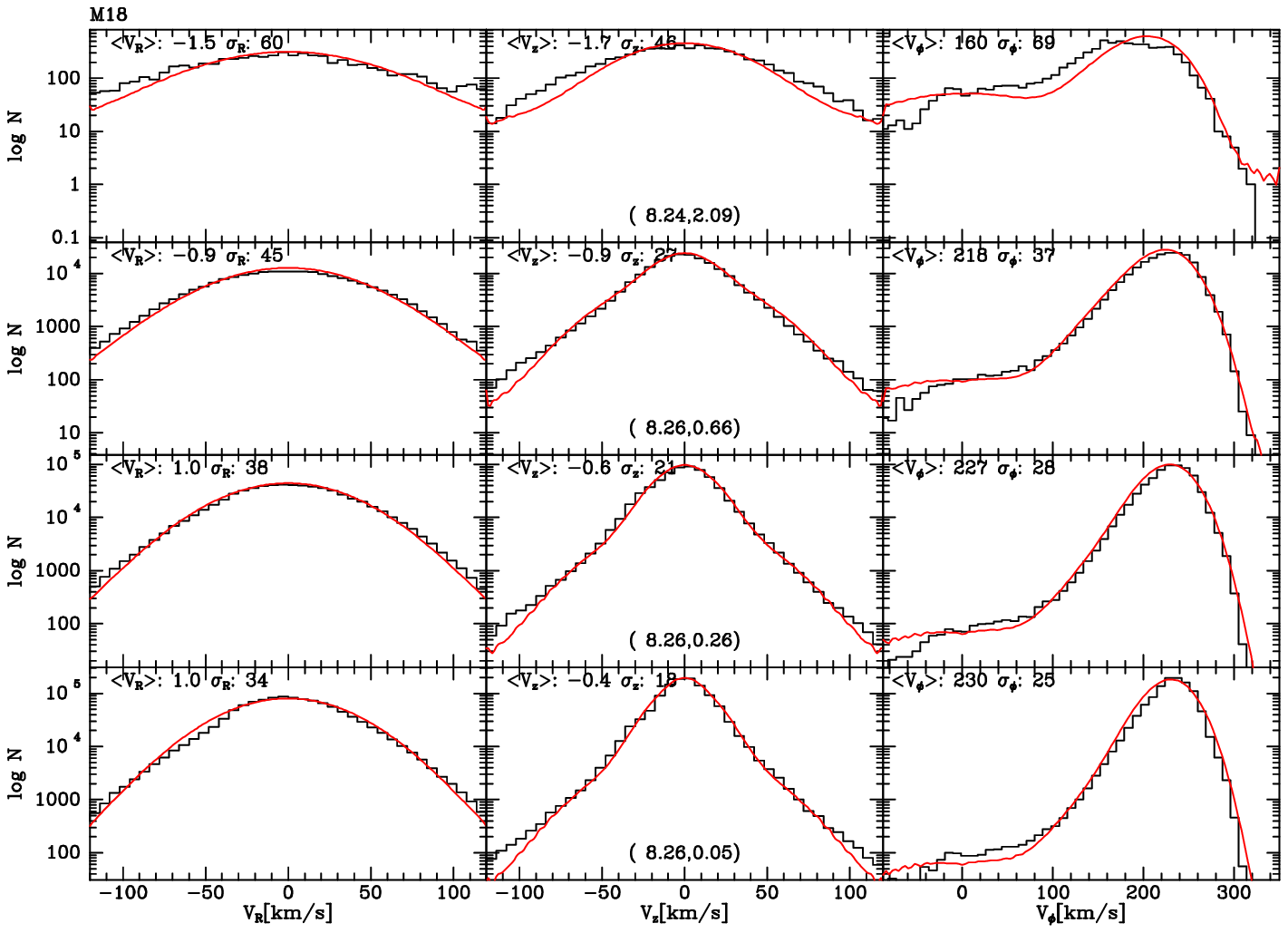}}
\centerline{\includegraphics[width=.8\hsize]{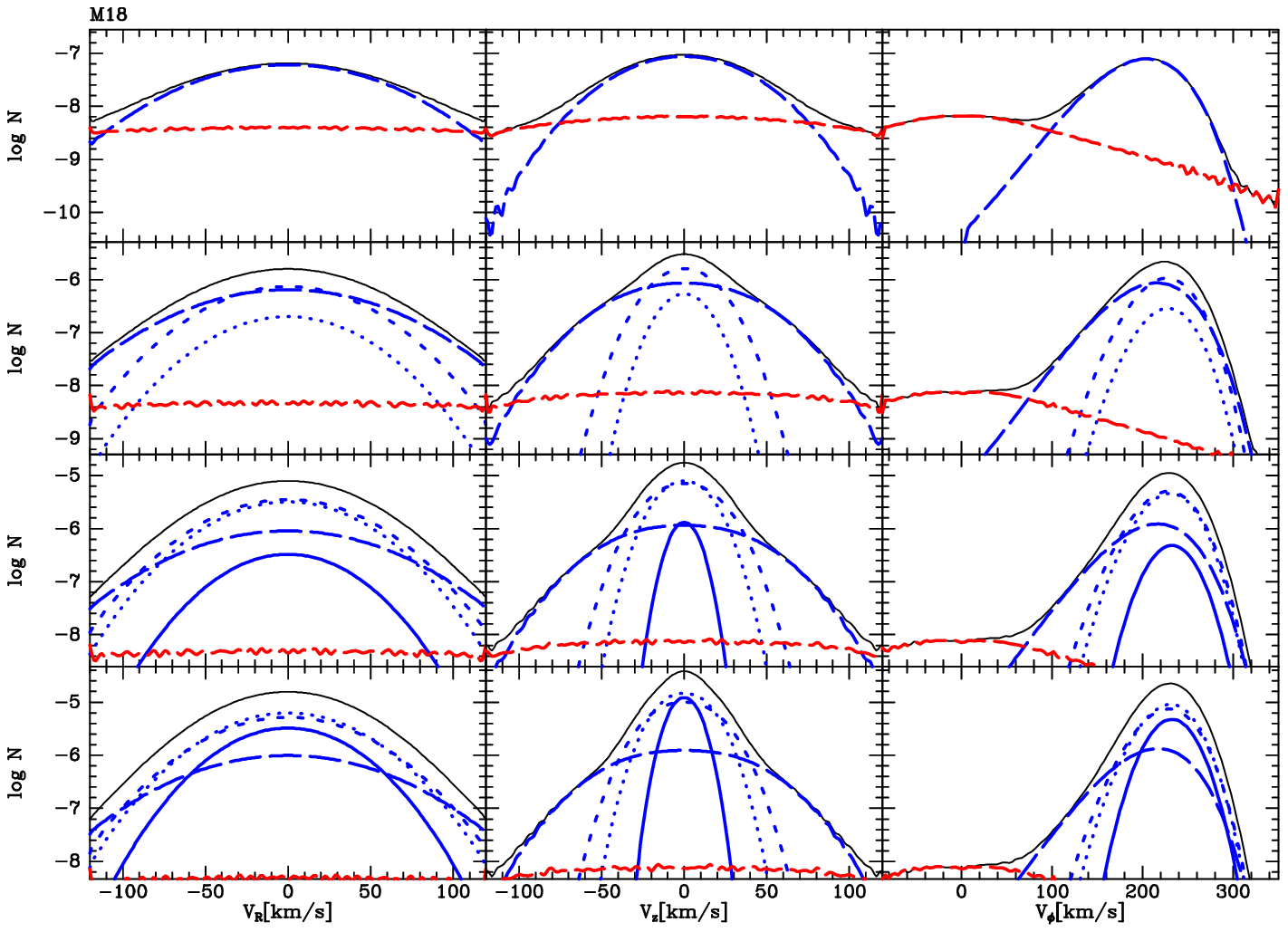}}
\caption{Upper block of panels: the same as Fig.~\ref{fig:Rin} but for
$R-R_0\in(-.5,+0.5)\kpc$.  Lower block of panels: how the model histograms
above are assembled from the contributions of different components. Blue
curves show the disc's contributions (full, dotted and dashed as age
increases and long-dashed for the thick disc). The contributions from the
bulge and the stellar halo are plotted in heavy black and red, respectively.
}\label{fig:R}
\end{figure*}

\begin{figure*}
\centerline{\includegraphics[width=.8\hsize]{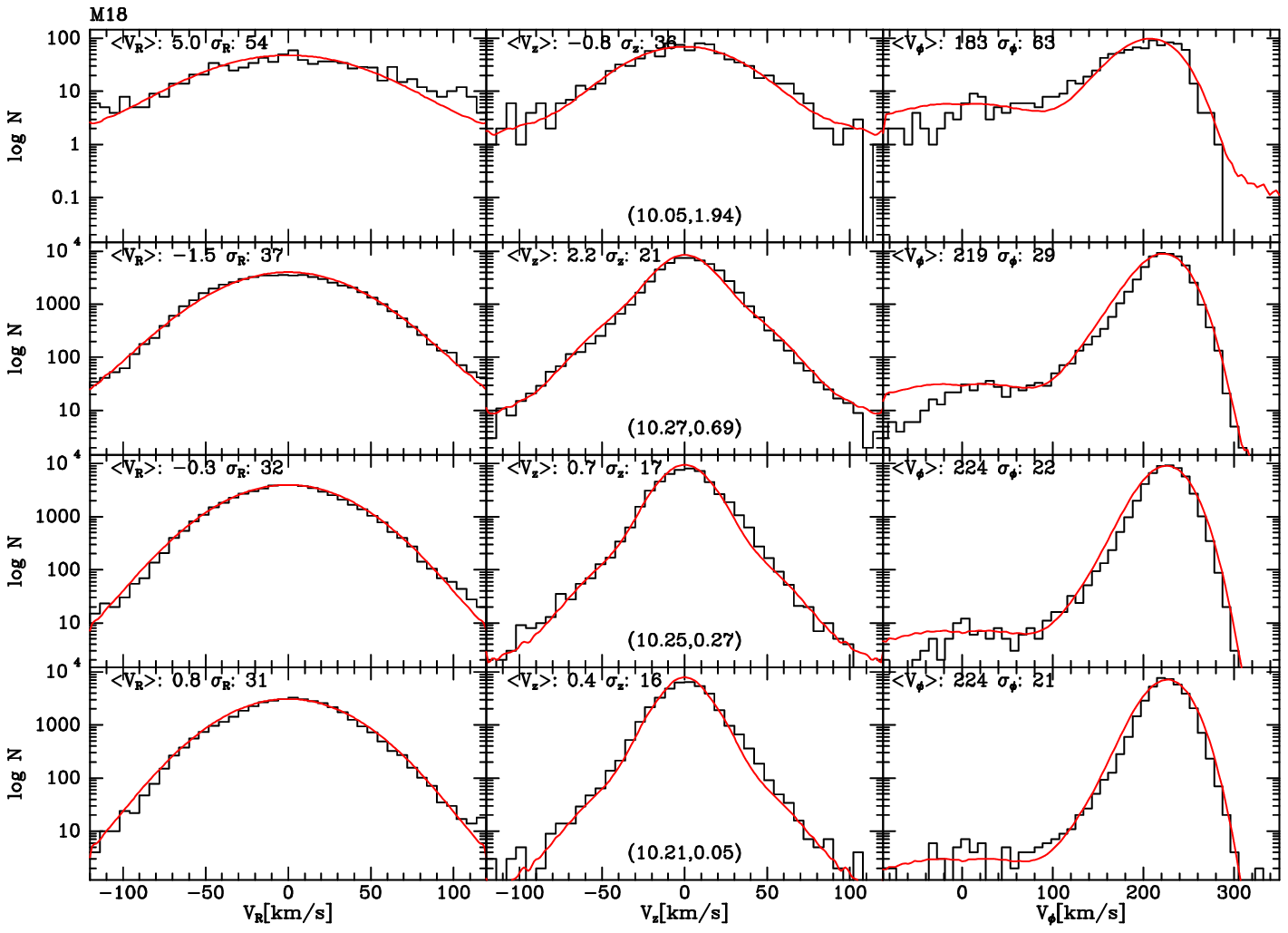}}
\caption{The same as Fig.~\ref{fig:Rin} but for  $R-R_0\in(1.5,3)\kpc$.}\label{fig:Rout}
\end{figure*}

\subsection{Fitting Gaia data}\label{sec:Gaia_fits}

\agama\ computes the density of each component on an appropriate grid
(spherical for the dark halo and cylindrical for the stellar components)
using an assumed potential, then it solves Poisson's equation for the
resulting potential and re-determines the density in this new potential.
This sequence of potential determinations is quite rapidly convergent.  An
excellent first guess at the potential is generally available from the
potential of the last model created, so on a good laptop a new model can be
computed in only $\sim5$ minutes. After computing the self-consistent
potential, \agama\ computes one-dimensional velocity distributions at 35
locations $(R,z)$ that are the barycentres of spatial bins into which the
Gaia stars have been grouped. In Figs.~\ref{fig:Rin} -- \ref{fig:Rout} these
locations $(R,z)$ are shown at the bottom of the $V_z$ panels --
Table~\ref{tab:bdys} gives the bin boundaries.  The black histograms in
Figs.~\ref{fig:Rin} -- \ref{fig:Rout} show the velocity distribution of RVS
sample stars. We show histograms for only 12 of the 35 bins, being alternate
bins in both $R$ and $|z|$. The top left corner of each panel shows the mean
and standard deviation of the histogram. The small, varying non-zero means of
$V_R$ and $V_z$ must arise from a combination of the Galactic bar and
non-equilibrium dynamics. The red curves in Figs.~\ref{fig:Rin} --
\ref{fig:Rout} show the velocity distributions computed by \agama, normalised
to the same number of stars as the corresponding black histogram.

\begin{figure}
\centerline{\includegraphics[width=\hsize]{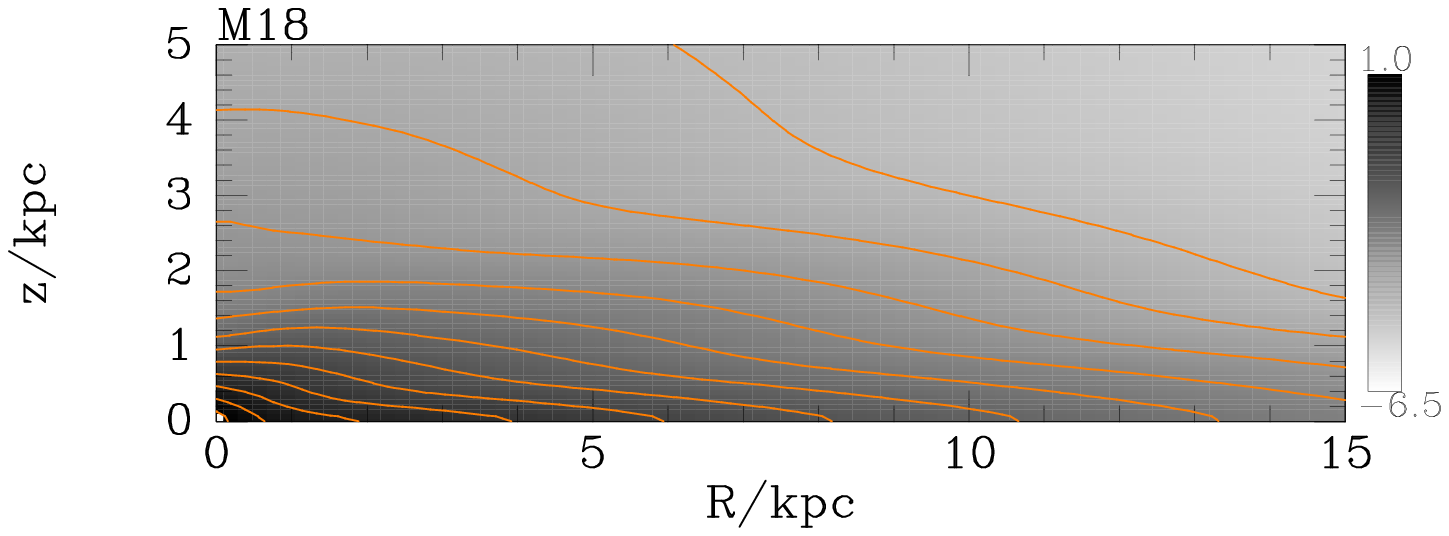}}
\centerline{\includegraphics[width=\hsize]{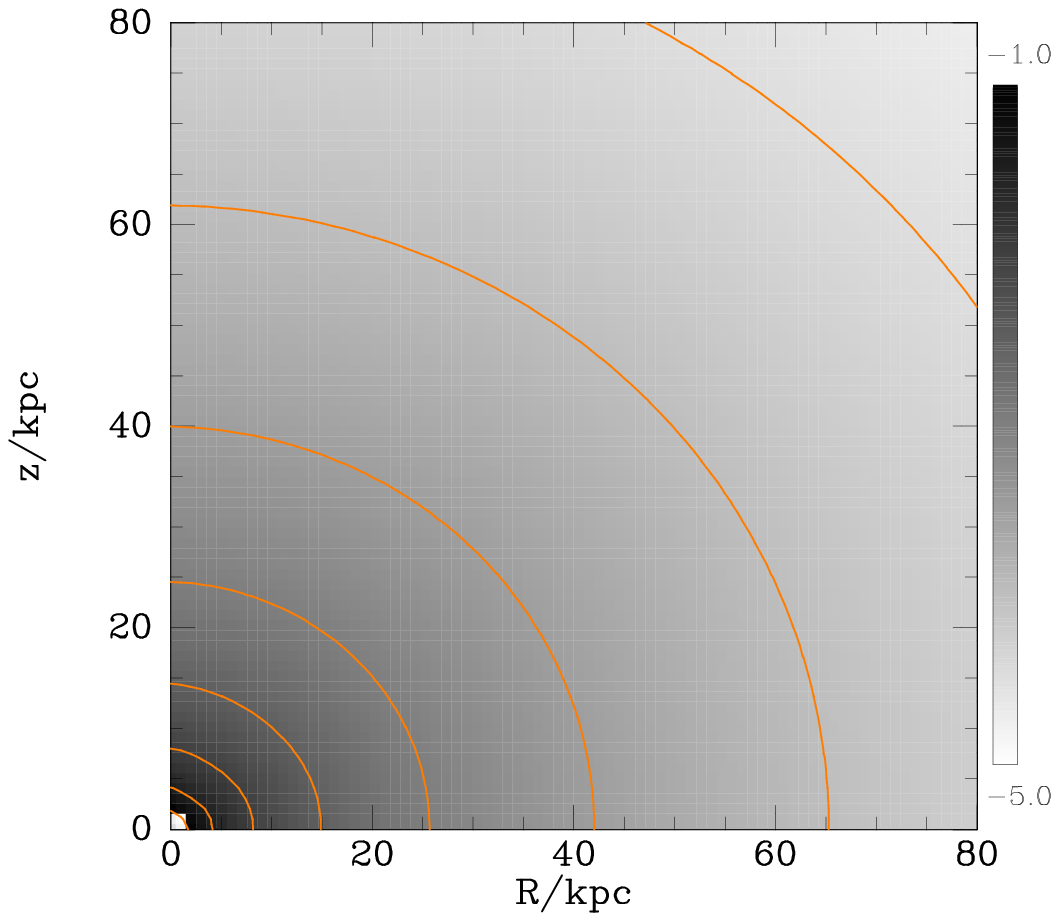}}
\caption{The density in the meridional $(R,z)$ plane of stars
(upper panel) and  dark matter. The colour scales are logarithmic with two
contours per dex.}\label{fig:merid_plots}
\end{figure}

Experiments with automatic optimisation of the parameters by the Nelder-Mead
downhill simplex algorithm \citep[e.g.][]{NumRec} were unsuccessful in that the machine proved
unable to find a convincing DF in an acceptable number of iterations. It
seems that the high dimension of the relevant parameter space combined with
the cost of each model evaluation (which arises more from computation of
diagnostics than from model construction) sets a requirement for a more
sophisticated machine-learning algorithm than naive least-squares
minimisation. Therefore we now present a model that has been fitted by hand
to the Gaia data and the vertical stellar density profile of \cite{GiRe83}.
Table~\ref{tab:DF_G} lists the values of the parameters that define the
model. Fig.~\ref{fig:merid_plots} shows the density of the stars and dark
halo in the meridional plane.

The lower block of panels in Fig.~\ref{fig:R} breaks the model velocity
histograms for the solar cylinder into the contributions of each stellar
component. The full blue curve of the young disc only dominates at the core
of the $V_z$ distribution, while the thick disc dominates in the wings and
furthest from the plane everywhere except in the $V_\phi$ histogram, where at
small $V_\phi$ the stellar halo dominates. The values of $J_{z0}$ listed in
Table~\ref{tab:DF_G} show a six-fold increase in $J_{z0}$ between the thick
disc and the old disc, whereas the characteristic actions $J_{r0}$ that
control the in-plane dispersions increase by a factor of six only between the
young disc and the thick disc. Another noteworthy trend among the parameters
are the steady increases in the values of $p_r$ and $p_z$ as one passes from
the young disc to the thick disc. That is, the decrease  outwards of the
velocity dispersions is steeper in older components. 

When fitting models by hand, velocity histograms like those shown in
Figs.~\ref{fig:Rin} to \ref{fig:Rout} yield numerous clues for parameter
improvement. Most fundamentally, the location of the bumps in the $V_\phi$
histograms at bottom right indicate how good the model's circular-speed
profile is. This informs the choice of the normalisations of the principal
components (dark halo, bulge, discs). The widths of the $V_R$ distributions
shown in the left columns of Figs.~\ref{fig:Rin} to \ref{fig:Rout} guide
choices of the parameters $J_{r0}$ of the disc DFs, with the youngest disc
dominating the histograms for low $|z|$ (bottom left panels) and the thick
disc dominating the histograms at top left. Any tendency for the model $V_R$
histograms to be less satisfactory at an inner radius than at an outer one is
addressed by adjusting the relevant $p_r$ parameter. Similarly, the $V_z$
histograms guide choices of $J_{z0}$, and $p_z$. Well above the plane, the
model $V_\phi$ histograms have a flat section at low $V_\phi$ and a bump.
The lower block of panels in Fig.~\ref{fig:R} shows that the flat section is
contributed by the stellar halo, while the bump is dominated by the thick
disc. The balance between these two features guides choices of the
normalisation of the stellar halo.

\begin{figure}
\centerline{\includegraphics[width=.95\hsize]{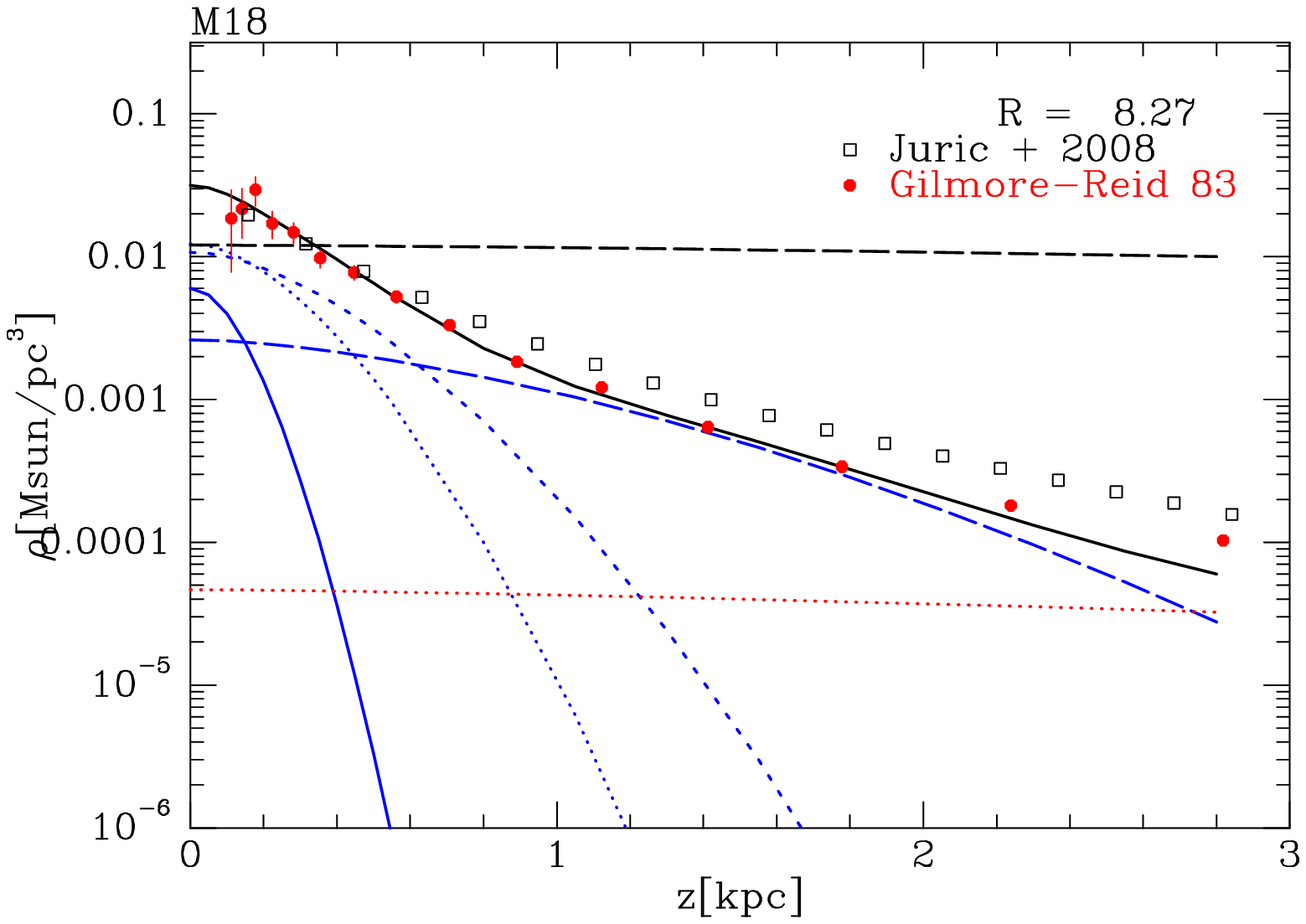}\quad}
\caption{Density as a function of distance from the plane at
$R_0$. From lower left to upper right, the blue curves show the contributions of the
young, middle-aged, old and thick discs. The dotted red curve shows the
contribution of the stellar halo and 
the full black curve shows the sum of
stellar contributions. The density of dark matter is shown by the dashed
black line. The red data points show the measurements of Gilmore \& Reid
(1983) while the open squares show the analytic fit to data of Juri\'c et
al.~(2008)}\label{fig:rhoz}
\end{figure}

\begin{figure}
\centerline{\includegraphics[width=.95\hsize]{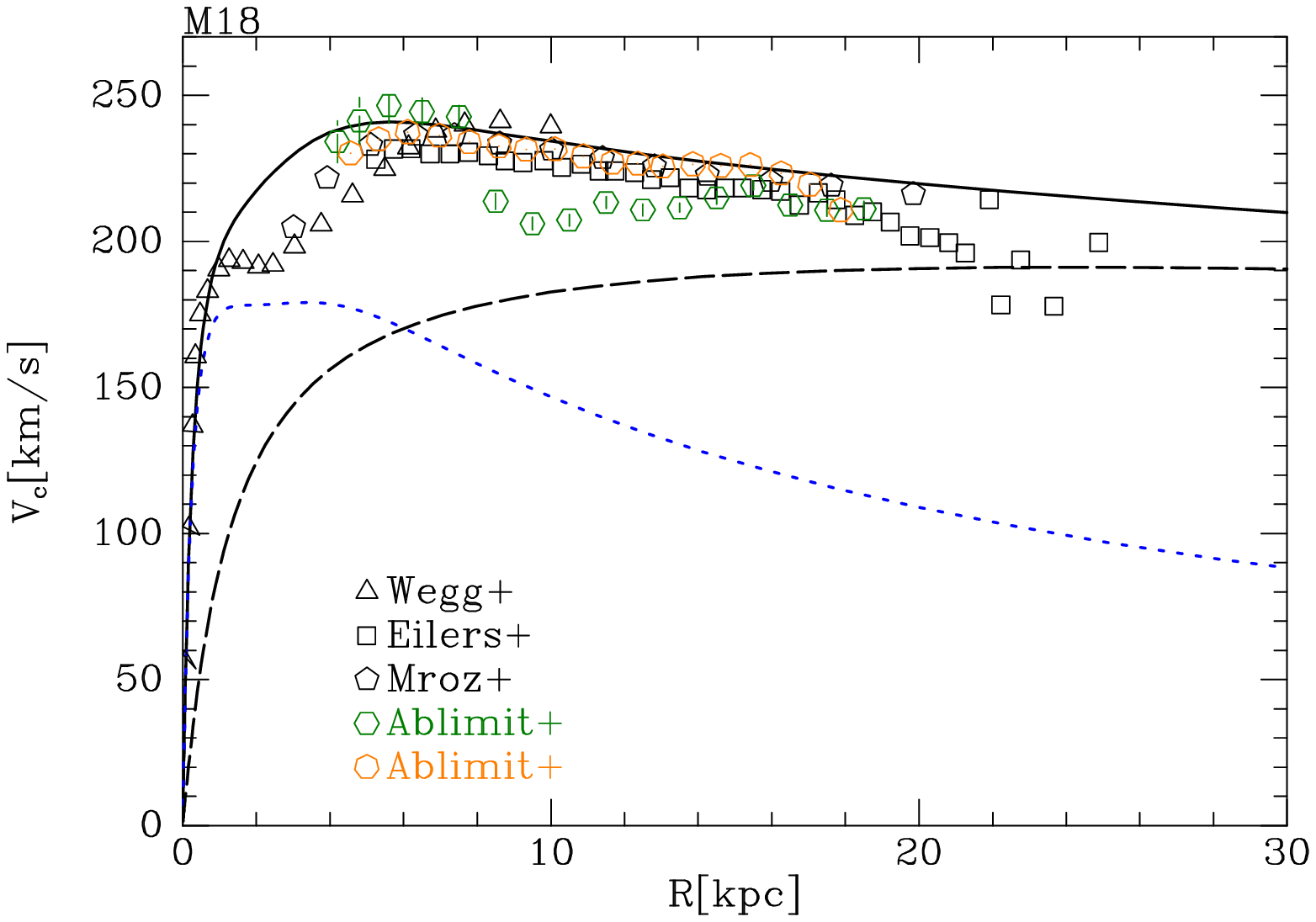}}
\caption{The circular speeds generated by the stars and gas (blue dotted)
and dark halo (black dashed) together with the total circular speed (full
black). The data points show estimates of $V_c$ from three recent studies.
The green circles show values obtained by Ablimit et al. (2020) using only
proper motions.
}\label{fig:Vc}
\end{figure}

Further clues to parameter choice can be drawn from Fig.~\ref{fig:rhoz},
which shows, in blue, the densities contributed at different distances from
the plane by the disc components (full curve: young disc; dotted: middle-aged
disc; short-dashed: old thin disc; long-dashed: thick disc). The dashed black
and dotted red curves, show the contributions of the dark and stellar halos,
respectively.  The full black curve shows the sum of the stellar components,
which may be compared with the data shown by red dots and black squares. The
red points are from \cite{GiRe83}, while the black squares show the
double-exponential fit of \cite{Juea08}. Both sets of points can be shifted
vertically at will.  To make the full black curve pass through the red data
points, one has to adjust the normalisations of the dark halo and the discs,
and also the disc parameters $J_{z0}$, in the choice of which
Fig.~\ref{fig:R} provides additional guidance. Crucially, the middle panels
of Figs.~\ref{fig:Rin} to \ref{fig:Rout} effectively fix the $J_{z0}$
parameters, so the
only way to address the model density falling off too steeply with $z$ is to
weaken the gravitational field near the plane by shifting mass from the disc
to the dark halo. Hence, if the black squares from the SDSS survey were
trusted more than the work of \cite{GiRe83}, one would make the dark halo
more massive and the disc less massive.

Comparison of the red model predictions and the black Gaia histograms in
Figs.~\ref{fig:Rin} to \ref{fig:Rout} shows a considerable measure of agreement
between the model and the data at each of the 12 locations shown; plots for
the remaining 23 locations considered show a similar level of agreement. 

\subsubsection{Weaknesses of the fits}

In the top-centre panels of Figs.~\ref{fig:Rin} and \ref{fig:R}, the red model
curves are narrower than the black data curves. The only significant
contributors to these curves are the thick disc, which dominates at
$|V_z|<90\kms$, and the stellar halo.  This problem might be addressed by
reducing the thick disc's value of $p_z$.

In the bottom left panel of Fig.~\ref{fig:Rin} the red model curve lies above
the black data curve. The thick, middle and old discs all make significant
contributions to the wings of these curves. The problem could be addressed by
lowering $J_{r0}$ for the middle disc at the risk of spoiling agreement in
other panels.

In Figs.~\ref{fig:Rin} to \ref{fig:Rout} the red curves for $V_\phi$ have a
tendency to be too high at $V_\phi<0$. 
It would be unwise to take the model's predictions for stars
of low $V_\phi$ too seriously because stars that approach the Galactic centre
must be influenced by the Galactic bar, which is not included in the model.
However, the model's prediction of a surfeit of stars at $V_\phi<0$ is least
apparent in Fig.~\ref{fig:Rin} for the bins closest to the centre, where the
bar should be most influential. The red curves for $V_\phi$ in the top-right
panels of Figs.~\ref{fig:Rin} to \ref{fig:Rout} are dominated by the
stellar halo, which we have modelled as a non-rotating component because many
studies \citep[e.g.][and references therein]{Schoenrich2014} have concluded
that this is so.  So perhaps one should revisit the issue of
halo rotation in light of these Gaia data: the outer halo may be non-rotating
(or even counter-rotating \citep{Carollo2007}) but the inner halo may rotate
detectably. In any event, ours is a very basic  model of the stellar halo,
that cannot address the empirical chemodynamical dichotomy into fairly
isotropic, metal-poor and radially biased, metal-richer components
\citep{Helmi2018,Belokurov2018}. Such a
model must await resolution of the issue regarding choice of DFs discussed by
\cite{Binney_in_prep}.

\subsubsection{Predictions of the models}
\begin{table}
\caption{Densities in the MW model in $M_\odot\pc^{-3}$ or $M_\odot\pc^{-2}$
($1\GeVdens=0.0263\msun\pc^{-3}$).}\label{tab:MW}
\begin{center}
\begin{tabular}{lcccc}
&gas&stars&DM&Total\\
\hline
$\Sigma(R_0,1.1\kpc)$&13.6&21.6&26.0&61.2\\
$\rho(R_0,0)$&0.057&0.0317&0.0121&0.0436\\
\end{tabular}
\end{center}
\end{table}

The full black curve in Fig.~\ref{fig:Vc} shows the circular-speed curve of
the gravitational potential that emerges from the choices made for the DF by
the self-consistency condition. After optimising agreement between model and
data in the earlier figures, the model does not have freedom to adjust
this curve, so it is a significant result that it agrees remarkably well with
the data points from other recent work, including the \cite{Wegg2013}
dynamical model of the bar/bulge, modelling of Cepheid variables by
\cite{Mroz2019} and \cite{Ablimit2020} and a
study of  $23\,000$ red giants by
\cite{Eilers2019}. In Fig.~\ref{fig:Vc} the model curve is set by the kinematics of all
stars in the restricted radial range $6\la R\la12\kpc$ rather than by objects
presumed to be on near circular orbits over a wider range of radii. The curve
extends in to the centre and out to large radii by virtue of the physical
assumptions that underlie the functional forms of the DFs. It
runs just above the majority of the data points,
but this will to some extent reflect our adoption from
\cite{SchoenrichME2019} of a faster value for the Sun's $V_\phi$ component
than others have used.

In Fig.~\ref{fig:Vc} the dotted blue curve shows the contributions to $V_c$
from stars and gas.  Baryons make the largest contribution to the
circular-speed curve within $6\kpc$ and the bulge contributes more than the
disc inside $2.3\kpc$. In fact, in the inner kiloparsec the gravitational
force from the disc is directed {\it outwards}, so it actually reduces $V_c$
-- see below.

Table \ref{tab:MW} gives the local densities of baryons and dark matter
according to the model shown here, and also the surface density
$\Sigma(R_0,1.1)=61.2\msun\pc^{-2}$ of mass that lies within $1.1\kpc$ of the
plane at the solar radius.  This comprises $13.6\msun\pc^{-2}$ of gas,
$21.6\msun\pc^{-2}$ of stars and $26.0\msun\pc^{-2}$ of dark matter. For
comparison, \cite{KuGi91}, using a sample of K dwarfs, inferred
$\Sigma(R_0,1.1)=71\pm6\msun\pc^{-2}$, while \cite{HolmbergFlynn}, using K
giants as tracers, obtained $\Sigma(R_0,1.1)=74\pm6\msun\pc^{-2}$. More
recently, by counting stars in Gaia EDR3 \cite{Everall_zb} infer
$(23\pm2.4)\msun\pc^{-2}$ of stars, and by applying the Jeans equations to
data from Gaia DR3 \cite{Nitschai2021} find
$\Sigma(R_0,1.1)=(55\pm1.7)\msun\pc^{-2}$. 

Table~\ref{tab:MW} reports the model's local density of stars to be
$0.0317\msun\pc^{-3}$, which lies below but is consistent with the value
$(0.0366\pm0.005)\msun\pc^{-3}$ \cite{Everall_zb} inferred from Gaia EDR3.
The local density of the stellar halo is $4.7\times10^{-5}\msun\pc^{-3}$.
Table~\ref{tab:MW} reports the local density of dark matter to be $\rho_{\rm
DM}(R_0)=0.0121\msun\pc^{-3}=0.46\GeVdens$.  For comparison, from their study
of red giants, \cite{Eilers2019} inferred the local density of dark matter to
be $0.0079\pm0.0008\msun\pc^{-3}$, while \cite{Nitschai2021} obtained
$0.0089\pm0.0006\msun\pc^{-3}$ and \cite{Cautun2020} found
$0.0088\pm0.0005\msun\pc^{-3}$. Thus our model is more dark-matter dominated
than several recent studies prefer. Evidently we still have some way to go before
a consistent picture emerges of the local structure of our Galaxy. A
comprehensive review of estimates of the dark halo can be found in
\cite{deSalas2021}. 

Given that the circular speed is falling at $R_0$, there is scope for
confusion in this area because the surface density $\Sigma(z)$ within $z$
of the plane and the vertical acceleration $K_z(z)$ satisfy
\[
\Sigma(z)={1\over2\pi G}\Big(K_z+{z\over R}{\d V_c^2\over\d R}\Big).
\]
Often values of   $K_z/(2\pi G)$ are quoted instead of $\Sigma$. In the
present model $K_z(1.1)/(2\pi G)=64.1\msun\pc^{-2}$

\begin{figure}
\centerline{\includegraphics[width=.9\hsize]{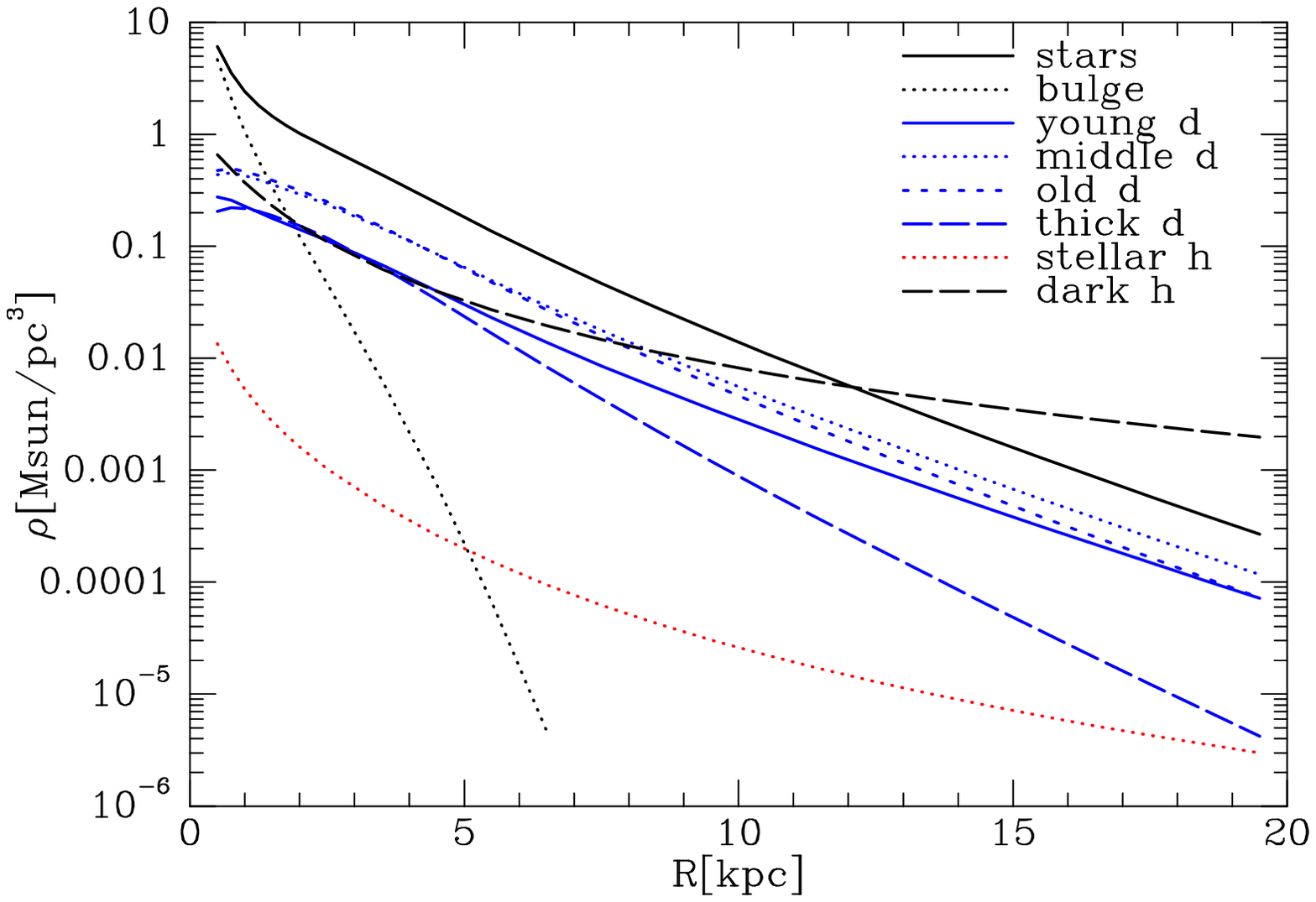}}
\centerline{\includegraphics[width=.9\hsize]{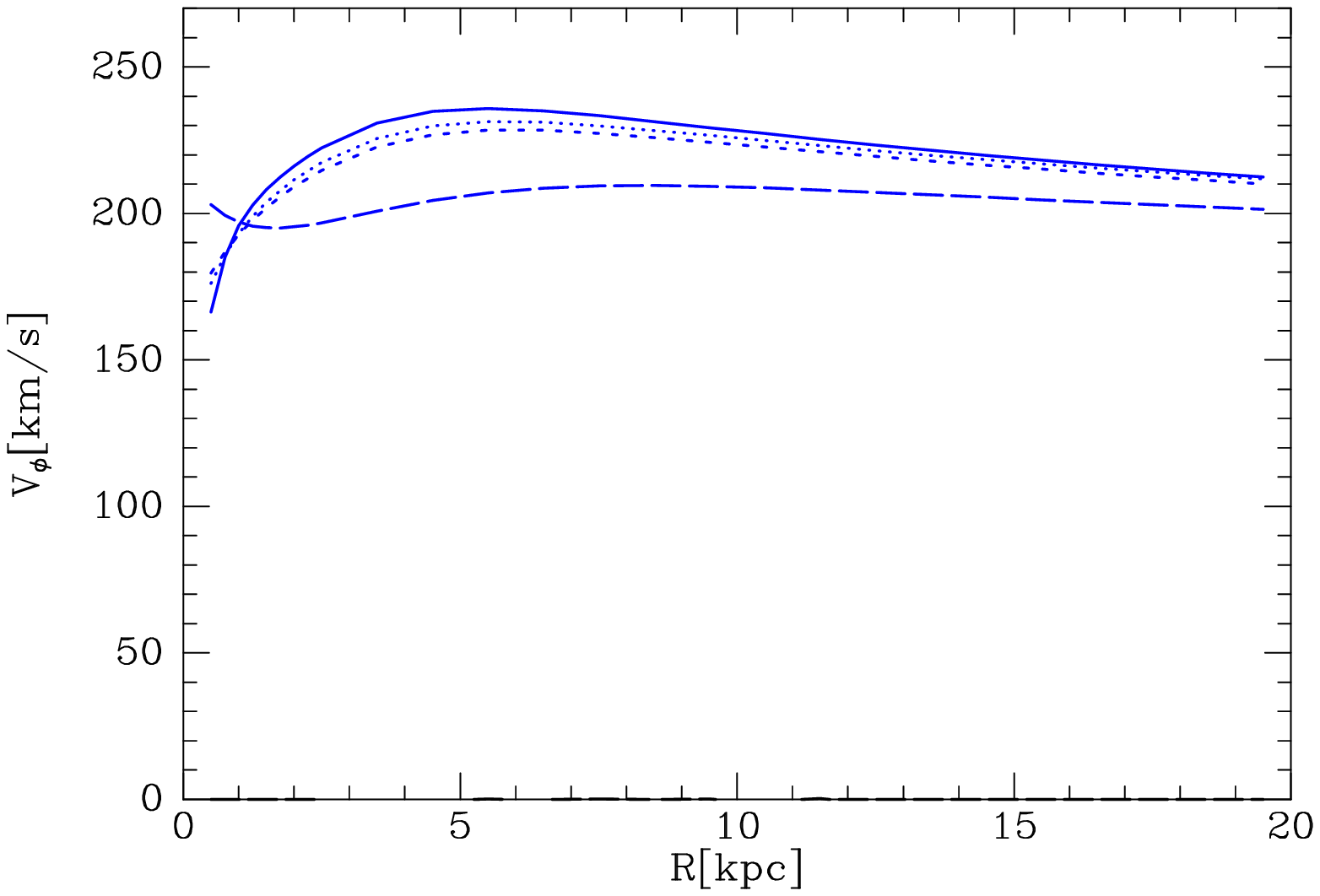}}
\caption{Upper panel: mid-plane densities of each component. Lower panel: mean
streaming velocities of disc components. Full black curves are for the young disc, dotted
curves are for the middle-aged disc, short-dashed curves are for the old disc and
long-dashed curves are for the thick disc. The dotted red curve in the upper panel
shows the density of the stellar halo and the full black curve shows the total stellar
density. The black long-dashed curves show the densities of the bulge and the dark
halo.}\label{fig:rhoR}
\end{figure}

\begin{figure*}
\centerline{\includegraphics[width=.3\hsize]{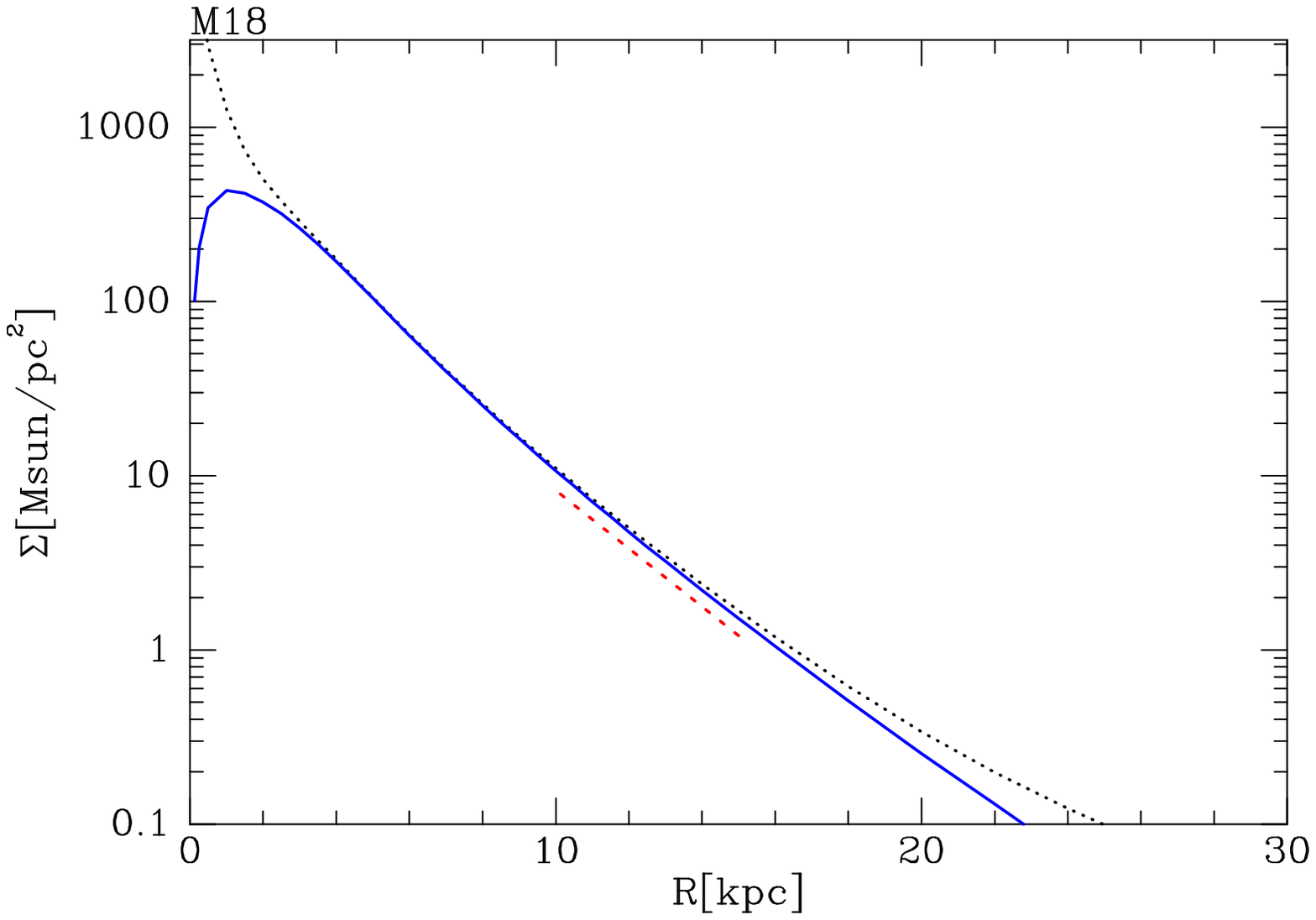}\quad
\includegraphics[width=.3\hsize]{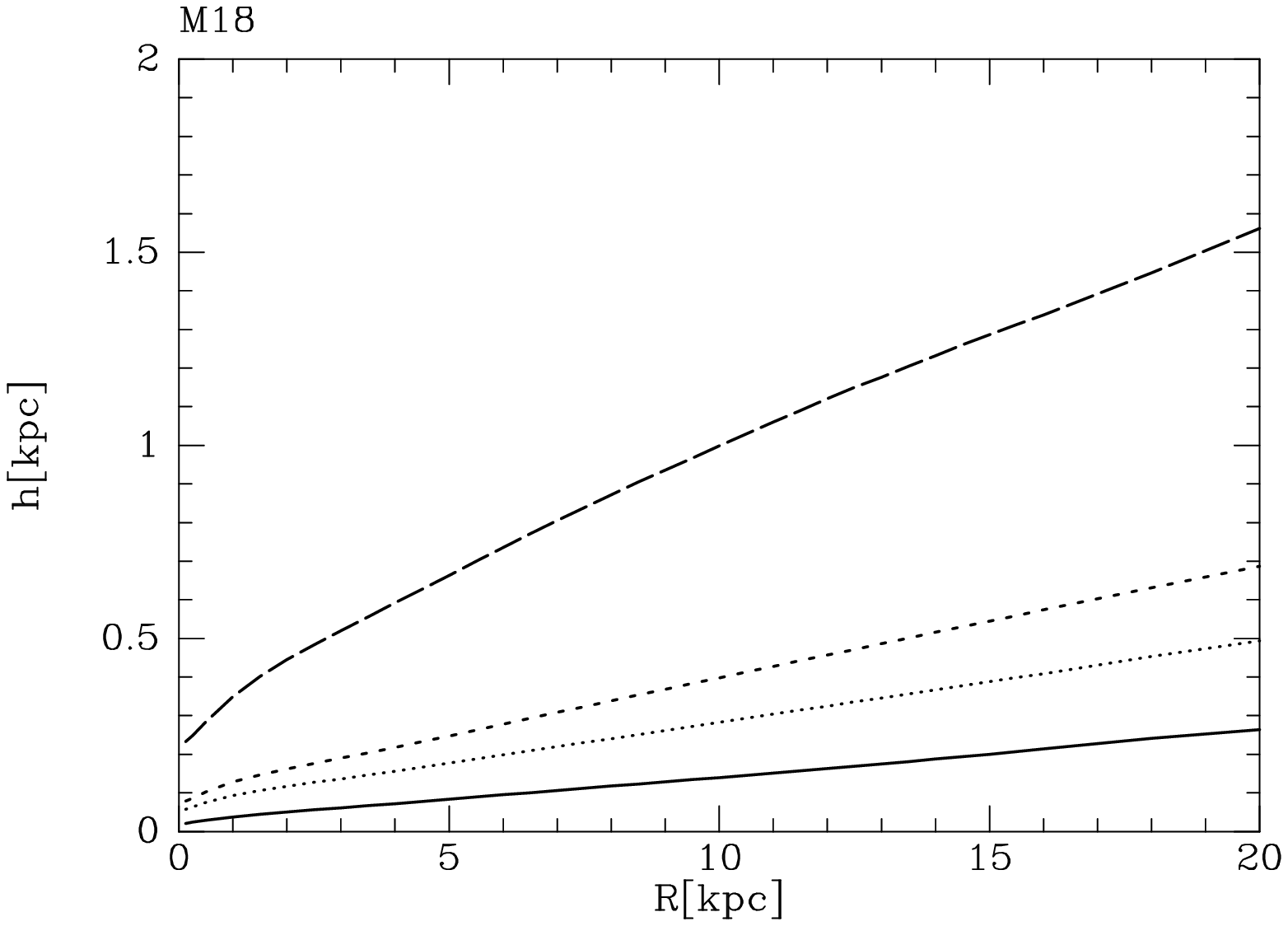}\quad
\includegraphics[width=.3\hsize]{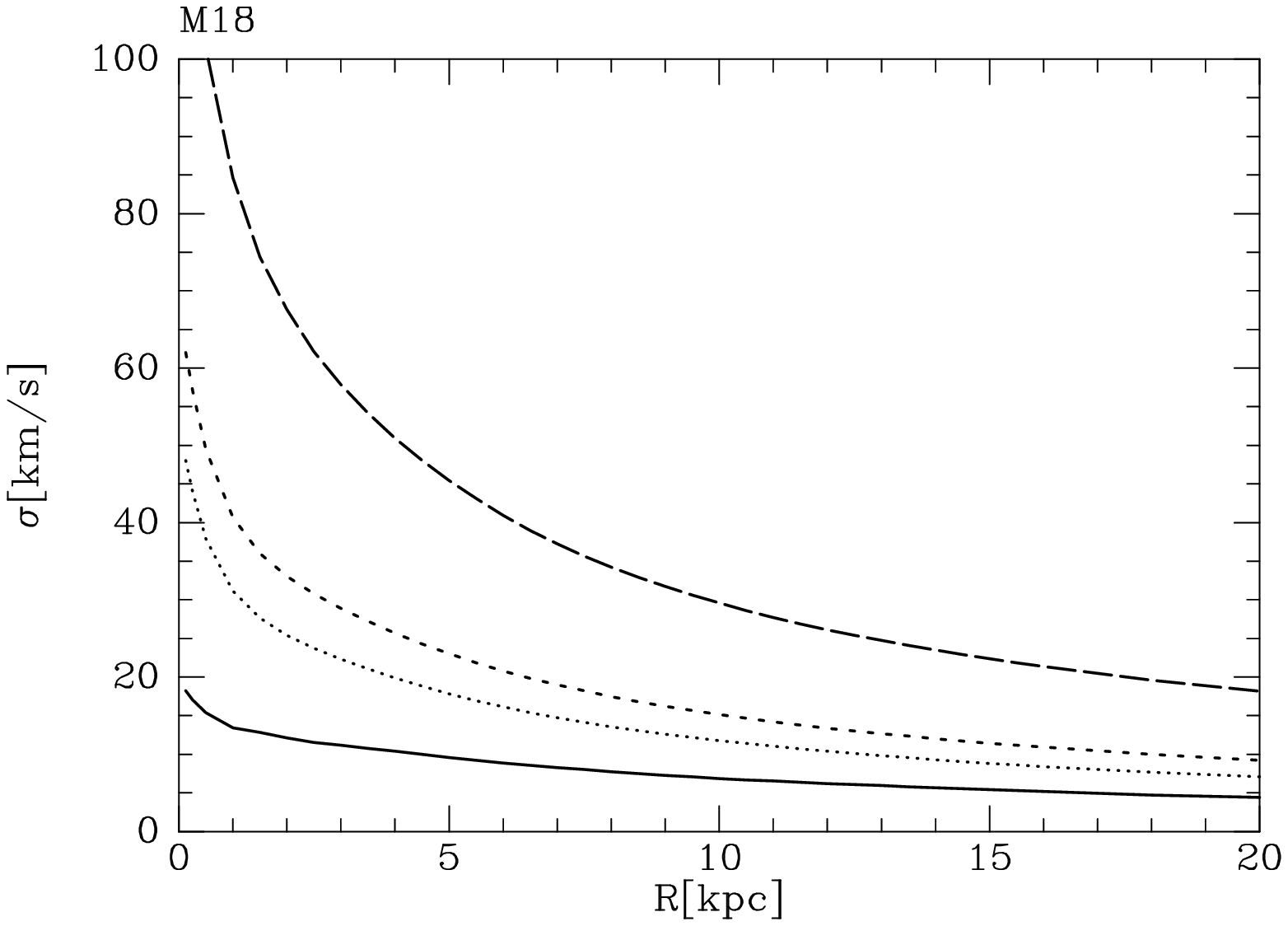}}
\caption{Quantities at face-on projection. 
Left: the disc's surface density (solid blue curve) and the surface
density including all stellar components (black dotted curve). The dashed red
line show the slope of a disc with scale length $2.6\kpc$. Centre:
rms heights of the disc components (full line for thin disc, dotted for
middle disc, short-dashed for old disc and long-dashed for thick disc). Right: projected line-of-sight
velocity dispersions of the disc components. }\label{fig:surfdens}
\end{figure*}

\begin{figure*}
\centerline{\includegraphics[width=.32\hsize]{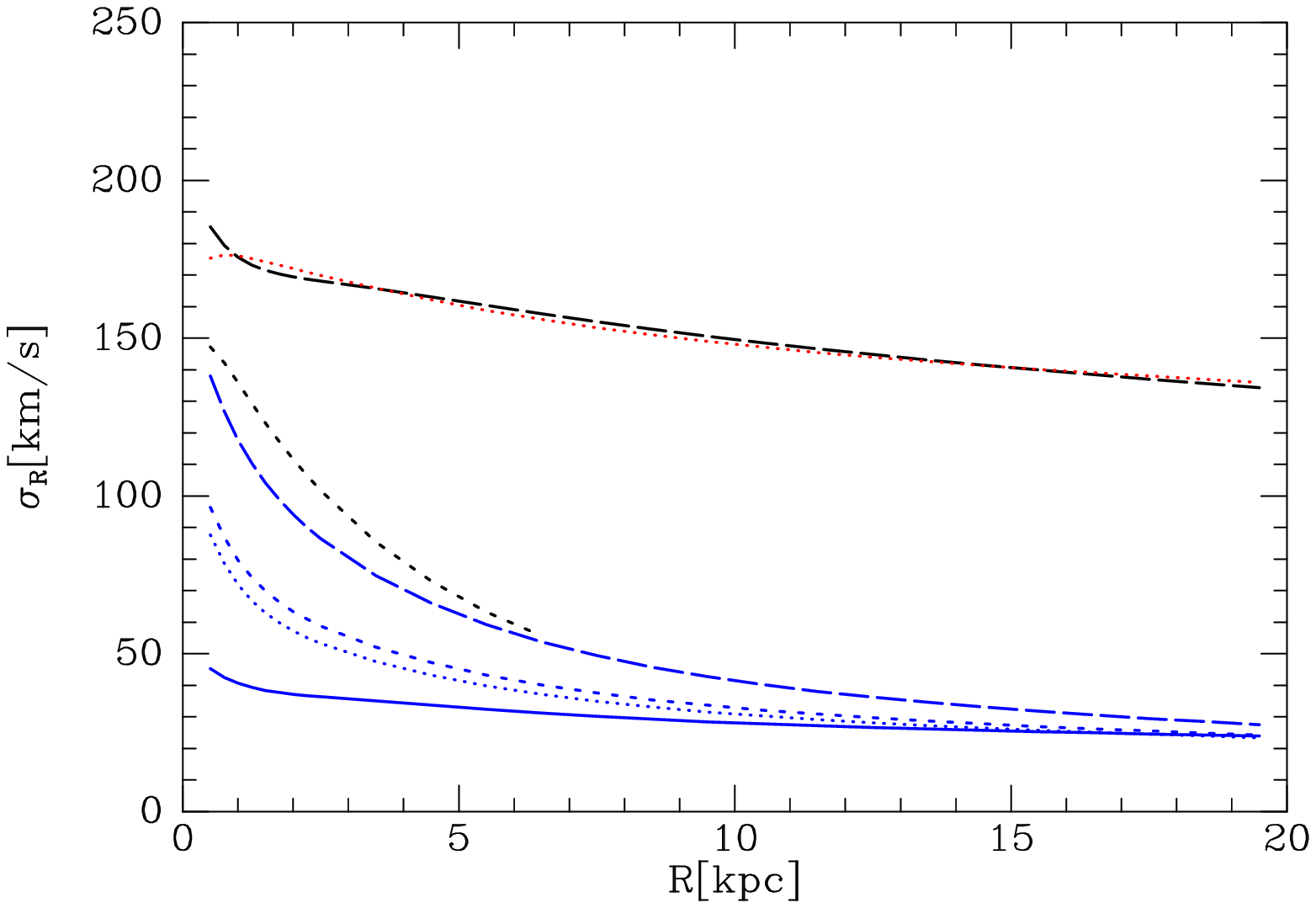}
\includegraphics[width=.32\hsize]{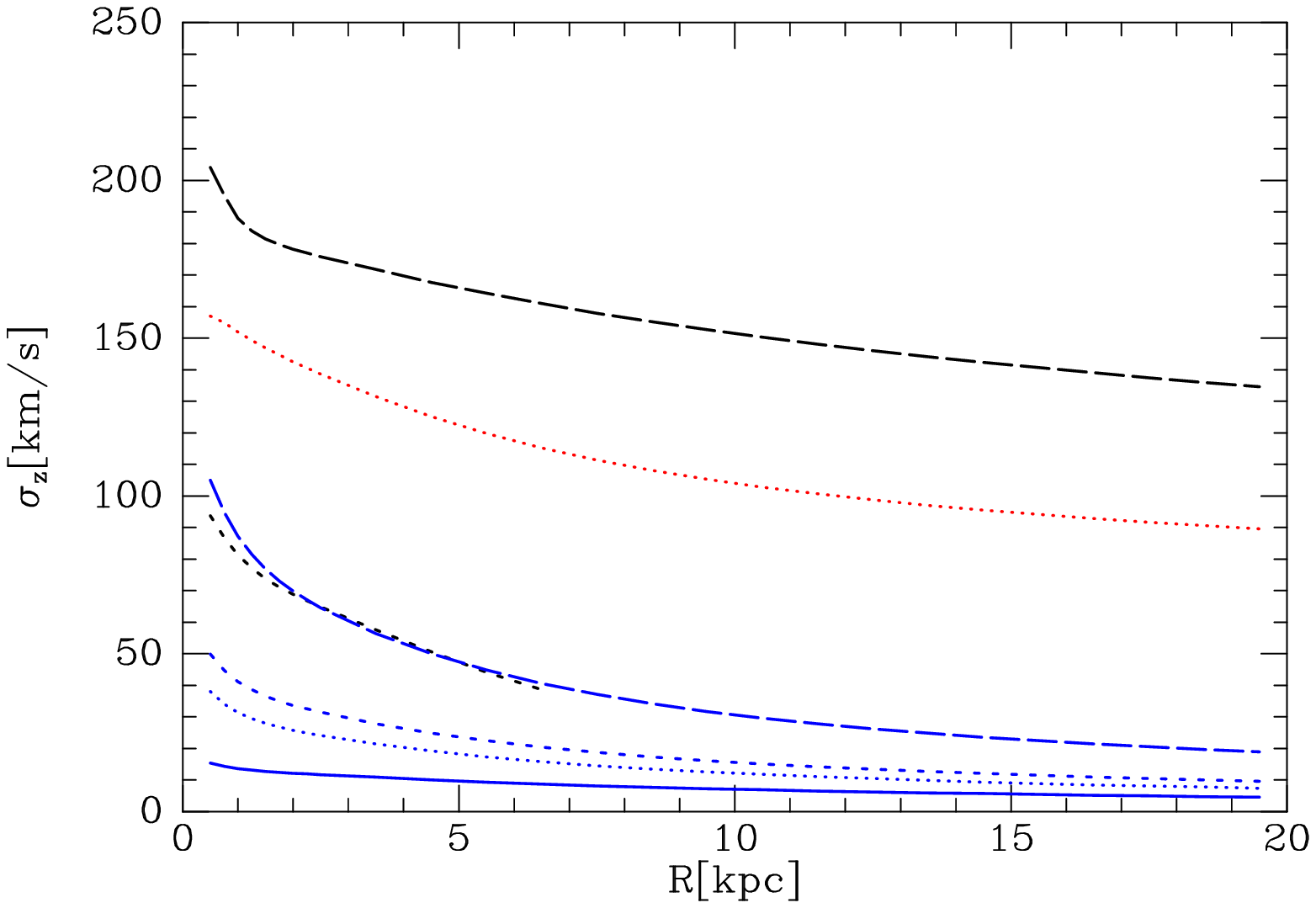}
\includegraphics[width=.32\hsize]{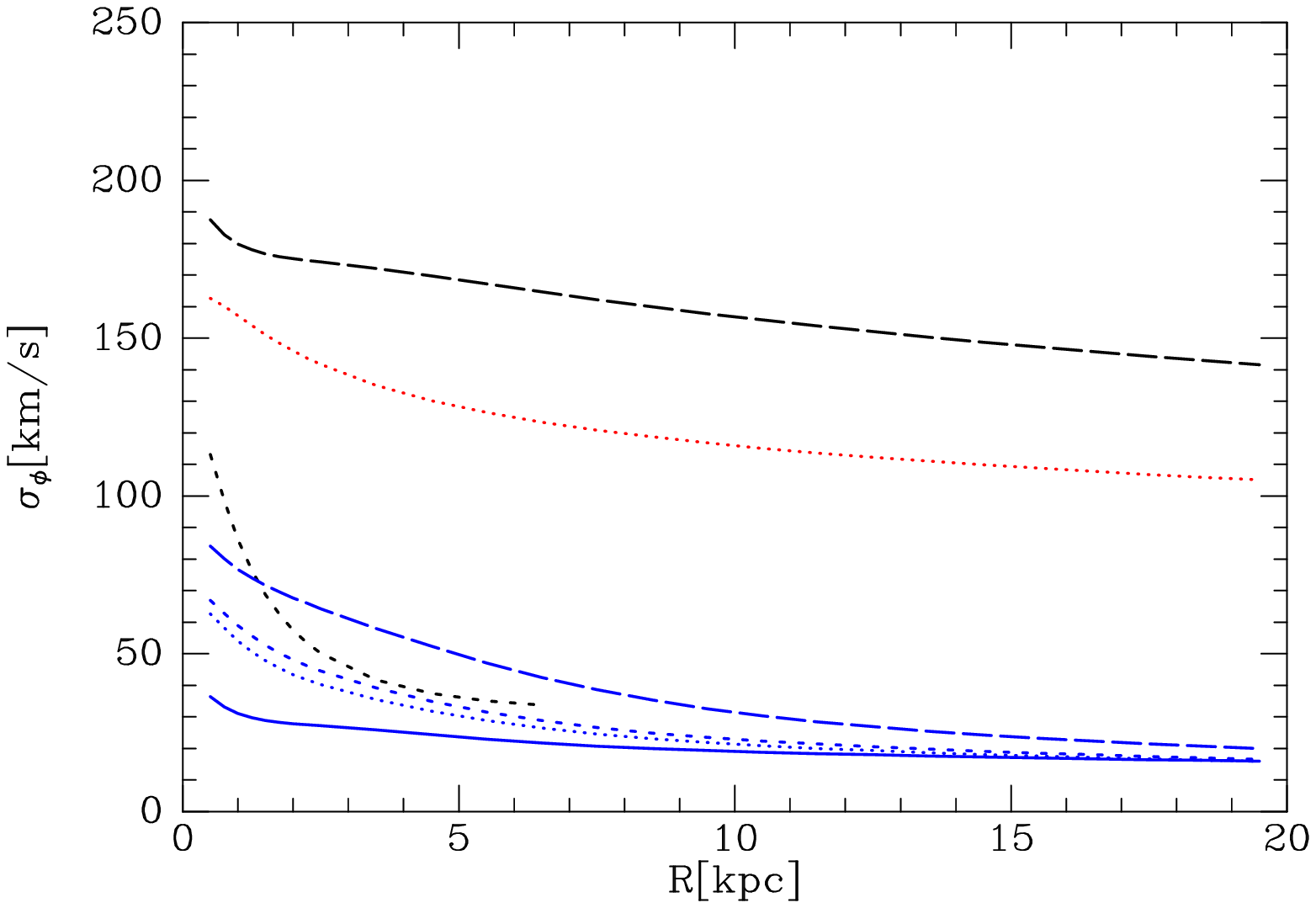}}
\caption{From left to right the radial dependence of the velocity dispersions $\sigma_R$,
$\sigma_z$ and $\sigma_\phi$. The black long-dashed curve is for the dark halo, the red
dotted curve is for the stellar halo. The dispersions of the discs are given by the four
blue curves at the bottom, with the young disc having smallest dispersions
and  the thick
disc the largest. The black dashed curves that reach only to $R=6\kpc$ show the dispersion in
the bulge.}\label{fig:sigR}
\end{figure*}

The full black curve in Fig.~\ref{fig:rhoR} shows the sum of the mid-plane
densities of all the stellar components. It is very nearly exponential
although comprised of contributions that have very different sale-lengths.
The steep black, dotted line shows the nearly exponential density of the
bulge. At the Sun the dark halo contributes a third of the density
contributed by stars.

The lower panel of Fig~\ref{fig:rhoR} shows the mean-streaming speeds
$\ex{V_\phi}$ of
the stellar discs at the mid-plane. For the thin-disc components
$\ex{V_\phi}$ is only slightly smaller than $V_c$, while the value for the
thick disc is generally reduced by an asymmetric drift in excess of $>10\kms$.

Fig.~\ref{fig:surfdens} shows what the Galaxy would look like when seen face-on.
The solid blue curve in the left panel shows the contributions of the four disc
components while the dotted black curve shows the density obtained on adding
the contributions of  the bulge and stellar halo. The profiles are not far
from exponential. 

The dashed red line shows the slope characteristic of a disc with a scale
length $R_\d=2.6\kpc$, slightly larger than the value $R_\d=2.4\kpc$ reported
by \cite{Roea03} from mid-IR photometry. There is a strong correlation
between the disc's scale length and degree of dark-matter domination in the
sense that longer scale lengths imply less dark matter.  This correlation is
enforced by the radial variation of the circular speed, which is very
tightly constrained by the $V_\phi$ histograms. If one assumes that the dark
halo is not too dissimilar to the NFW model (see Section
\ref{sec:no_baryons}), then the gentle decline in $V_c$ at $R\ga R_0$ evident
in Fig.~\ref{fig:Vc} places an upper limit on the dark halo's contribution
that becomes tighter as $R_\d$ grows, and, as explained at the end of
Section~\ref{sec:Gaia_fits}, a massive disc is incompatible with
the data for the vertical density profile shown in Fig.~\ref{fig:rhoz}. Thus,
ultimately the disc's vertical density profile at $R_0$ sets $R_\d$. 

The middle panel of Fig.~\ref{fig:surfdens} shows the rms thicknesses of the
four disc components. All four components become thicker as one moves
outwards. The right panel shows the projected line-of-sight velocity
dispersions of the disc components, which all increase inwards. The steep
central increases in the dispersions of the thin-disc components are
associated with the central holes in their surface densities: the
blue curves in left panel of Fig.~\ref{fig:surfdens} show that the surface
density of the stellar discs rises in the inner kiloparsec before flattening,
and starting to fall in earnest around $R=4\kpc$. This structure causes the
gravitational field of the stellar disc to be directed {\it outwards} near the
centre. The gas disc, which has a central hole, amplifies this effect.

Fig.~\ref{fig:sigR} shows the radial variation within the plane of the
velocity dispersions. At small radii the dark halo has a weak tangential
bias that is a consequence of its assumed adiabatic compression by the
baryons, and the dispersions fall
gently from  $>200\kms$ at $R\sim1\kpc$ to $120\kms$ at
$R=20\kpc$. The stellar halo (red dotted curves) is radially biased, with $\sigma_R$
similar to the dispersions of the dark halo.

The black dotted curves in Fig.~\ref{fig:sigR}, which show the velocity dispersions
of the bulge, show $\sigma_R$ and $\sigma_z$ tracking the dispersions of
the thick disc, but $\sigma_\phi$ falling much more steeply.

In all disc components $\sigma_R>\sigma_\phi>\sigma_z$.
everywhere although $\sigma_R$ invariably falls faster with increasing $R$
than does $\sigma_z$.

\subsection{The halo without baryons}\label{sec:no_baryons}

\begin{table}
\caption{Characteristics of the dark halo. Distances in kpc, masses in
$10^{10}\msun$.}\label{tab:Dhalo}
\begin{center}
\begin{tabular}{lccc}
&$r_s$&$r_{200}$&$M_{200}$\\
\hline
Dark halo&20.0&192&80.5\\
\end{tabular}
\end{center}
\end{table}

\begin{figure}
\centerline{\includegraphics[width=.8\hsize]{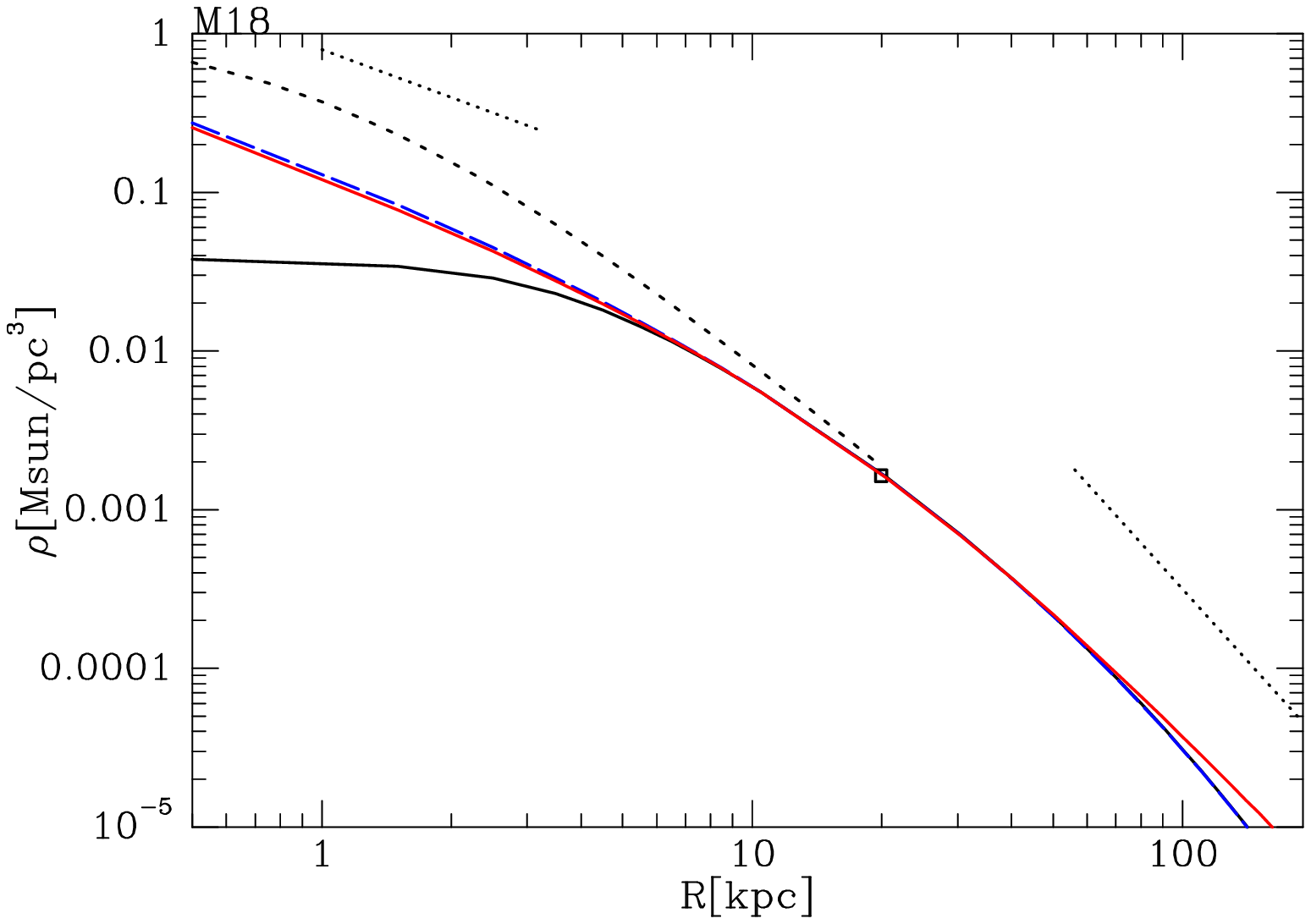}}
\centerline{\includegraphics[width=.8\hsize]{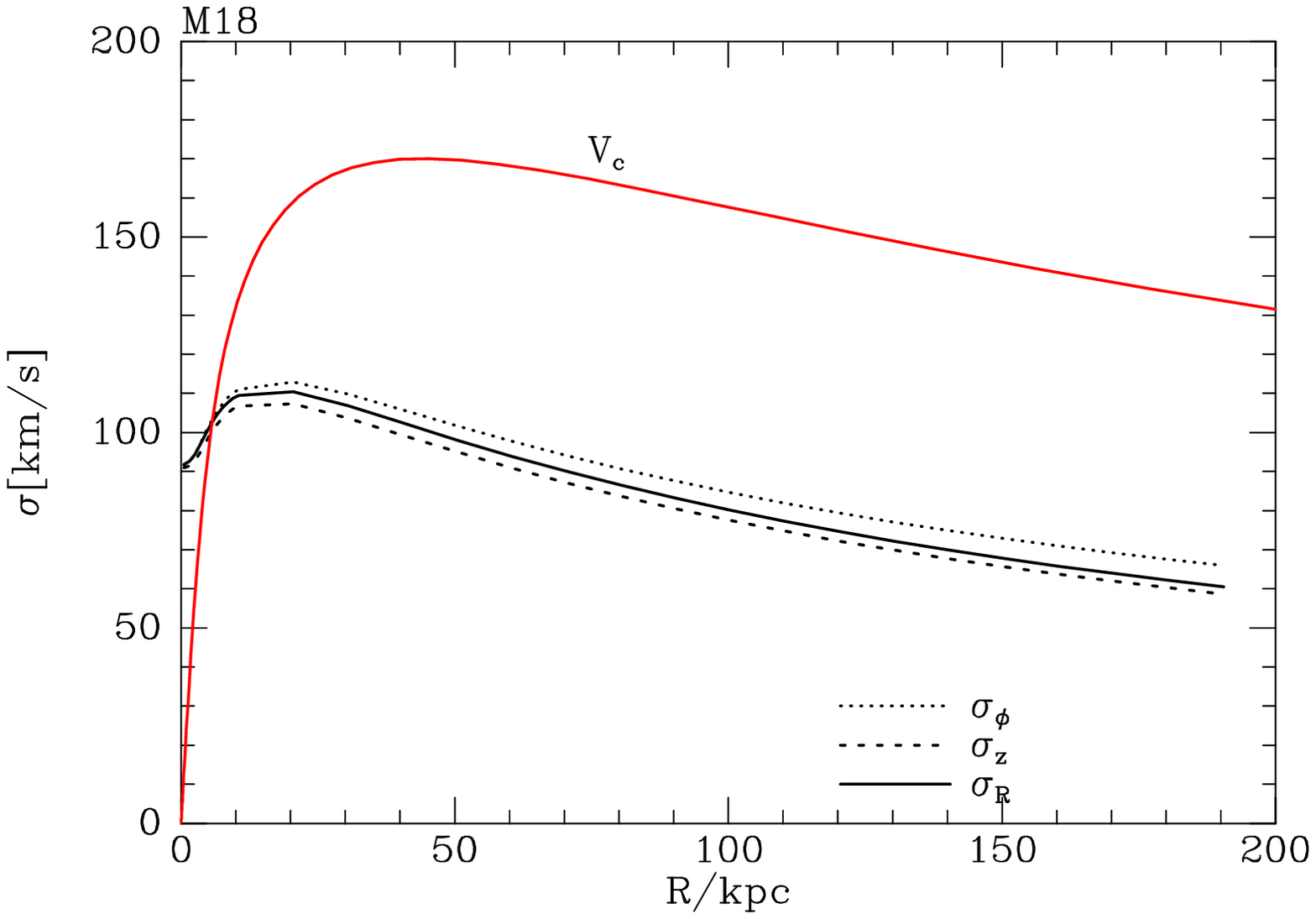}}
\caption{The structure of the dark halo before the infall of baryons. Upper
panel: the full black curve shows the density profile generated in isolation
by the DF of the Milky Way's dark halo. The dashed blue curve shows the
profile generated by this DF after removal of the Cole--Binney core. The
square marks the radius, $r_s=20.0\kpc$, at which $\d\ln\rho/\d\ln r=-2$ and
the red curve shows the density profile of the NFW halo with the same scale
radius. The black dashed curve shows the density of the dark halo as it
actually is in the Galaxy.  The sloping dotted
lines show the asymptotic slopes, $-1$ and $-3$ of an NFW profile.  Lower
panel: the circular speed and the principal velocity dispersions of the halo
before removal of the Cole-Binney core.}\label{fig:Dhalo}
\end{figure}

Our understanding of the statistical properties of dark haloes depend to a
large extent on cosmological simulations that exclude baryons. Comparison of
the Galaxy's dark halo with expectations raised by such simulations requires
knowledge of the structure of the dark halo before the baryons fell in. The
classic assumption is that the accumulation of baryons was an adiabatic
process \citep{Blea86,SeMcG05}. In this case, the primordial structure of the dark
halo can be recovered by constructing the self-consistent model defined by
the dark halo's DF alone.  The full black curve in
Fig.~\ref{fig:Dhalo} shows the structure of this object.

There is now significant evidence that baryonic infall was significantly
non-adiabatic \citep{Chan2015,PontzenGovernato2014,Pascale2018}. Indeed it is
natural that in the region in which baryons dominate the gravitational field,
fluctuations in the gravitational field will have upscattered dark-matter
particles, with the consequence that their phase-space density is now lower
than it was before the baryons fell in.  The Cole-Binney core of our halo is
designed to model the consequences of this physics, so to obtain the best
estimate of what the dark halo looked like before the baryons fell in, we
should remove this core from the DF. The dashed blue curve in
Fig.~\ref{fig:Dhalo} shows the density profile of this primordial dark halo.
It closely hugs the red curve, which plots the density of the NFW model that
has the same scale radius $r_s=20\kpc$ (the radius at which $\d\ln\rho/\d\ln
r=-2$). This closeness of fit is no accident: the search for a dark halo DF
was confined to ones that in isolation generate an NFW profile after removal
of the Cole-Binney core. This curve yields $r_{200}=192\kpc$ for the radius
at which the mean density is 200 times the mean cosmic density and
$M_{200}=0.805\times10^{12}\msun$ for the mass interior to that radius.  For
comparison, \cite{Eilers2019} derive $M_{200}=(0.725\pm0.026)\times
10^{12}\msun$, while two estimates obtained by \cite{Ablimit2020} are
$[r_s=(14.5\pm0.5)\kpc,\,
M_{200}=(0.66\pm0.07)\times10^{12}\msun,\,r_{200}=(179\pm5)\kpc]$ and
$[r_s=(14.7\pm0.4)\kpc,\,M_{200}=(0.82\pm0.05)\times10^{12}\msun,\,r_{200}=(192\pm4)\kpc]$.
When comparing these estimates one must bear in mind that the RVS data can
constrain the density only at $R\la12\kpc$, so quoted values of even $r_s$
involve significant extrapolation and values of $r_{200}$ and $M_{200}$ are
dangerously exposed to assumptions about the halo's density profile -- for
example, the mass of our halo could be made larger either by increasing its
truncation action $J_{\rm cut}$ or making its outer slope shallower. The
globular cluster system provides some sensitivity to mass that lies beyond
$r=12\kpc$, and from a study of this system \cite{Wang2021} inferred $0.54\la
M_{200}/10^{12}\msun\la0.78$. Using both globular clusters and dwarf
spheroidal galaxies, and taking into account perturbation by the LMC,
\cite{Vasiliev2022} finds $M_{200}=(0.7-1.6)\times10^{12}\msun$.  Whereas
ours estimates are based on a reconstruction of the dark halo prior to baryon
infall, those of earlier work relate to models of the halo as it now is.

The dashed curve in the upper panel of Fig.~\ref{fig:Dhalo} shows the dark
halo's current density profile; despite the upscattering of dark-matter
particles, the gravitational pull of the baryons has more than tripled its
density at $R\simeq1\kpc$ from what it was even before the
upscattering. The density enhancement remains significant out to
$r_s$. Comparing the current of the dashed and full black curves in the upper
panel of Fig.~\ref{fig:Dhalo} we see that the widespread assumption that the
dark haloes of luminous galaxies have NFW profiles is indefensible. Dark
haloes will
have such profiles only if upscattering of dark-matter particles precisely
cancels adiabatic compression, and there are no grounds for believing this to
be true. For more on this topic, see \cite{Cautun2020}.

The lower panel of Fig.~\ref{fig:Dhalo} plots velocities associated with this
reconstruction of the primordial dark halo. Its circular speed, plotted in
red, peaks at $170\kms$ and its velocity distribution is nearly isotropic.
Removal of the core changes the dispersions negligibly at $R\ga5\kpc$.

\subsection{Chemodynamics}\label{sec:Hayden}

\begin{figure*}
\centerline{\includegraphics[width=\hsize]{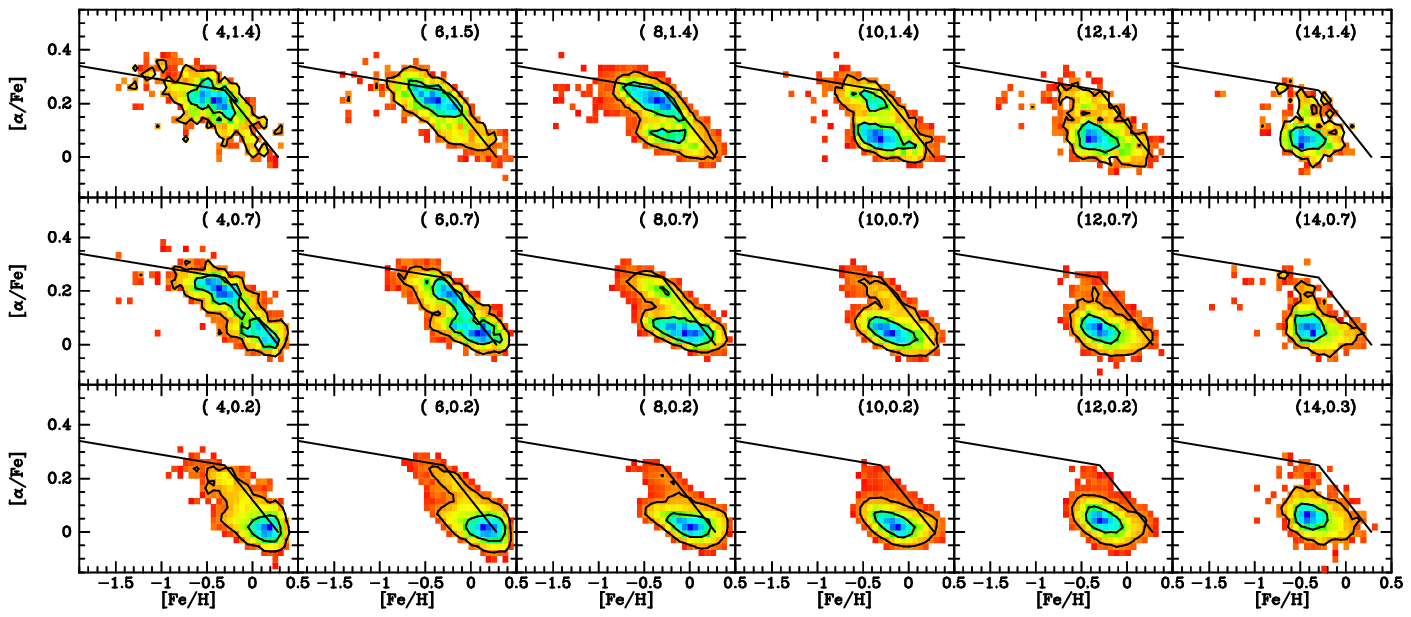}}
\medskip
\centerline{\includegraphics[width=\hsize]{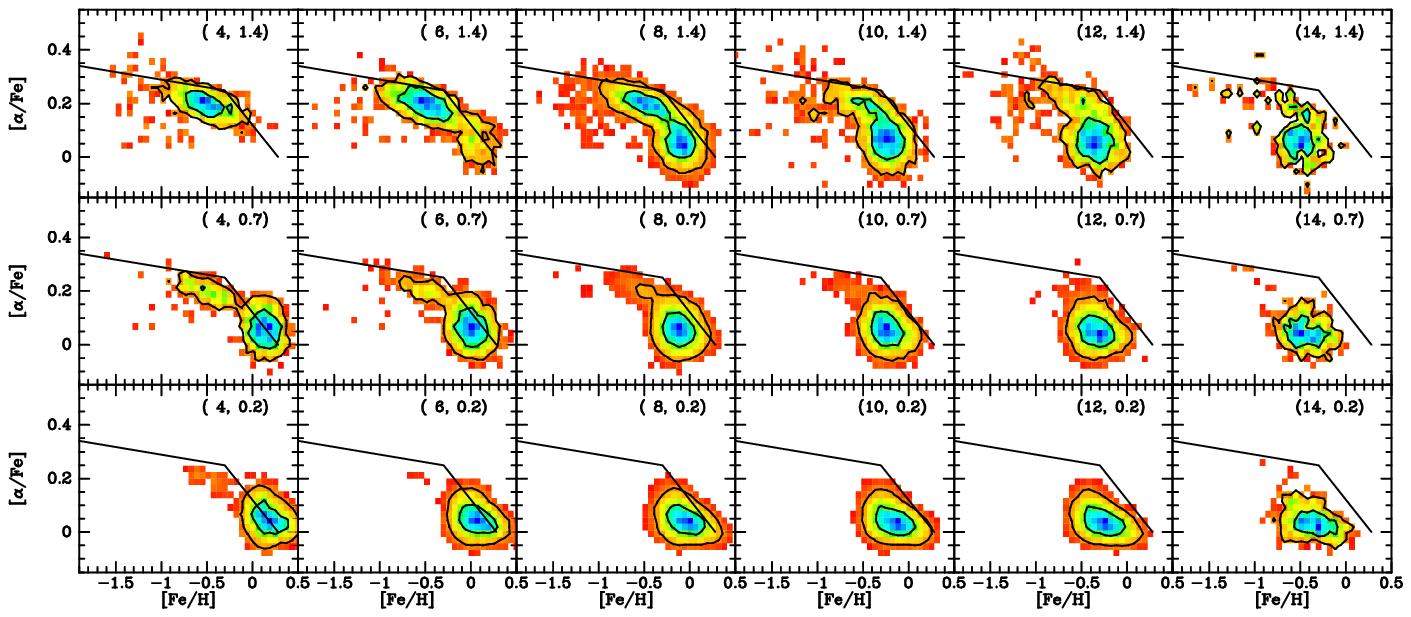}}
\caption{Upper block of panels: a plot analogous to  Fig.~4 in Hayden et al.\ (2015)
using the same APOGEE data. Lower block: a similar plot computed from our
dynamical model using the chemical
compositions of the components given by equations (\ref{eq:chemo_first}) to (\ref{eq:chemo_last}). 
The mean values of $R$ and $z$ for the stars/pseudo-stars that fall in
each spatial bin are given at upper right of each panel.
Contours are drawn that enclose 90 percent and 50
percent of stars.
}\label{fig:Hayden}
\end{figure*}

Fig.~4 of \cite{Hayden2015} is one of the most significant products of the
APOGEE survey \citep{Majewski2017}. It shows the distribution of stars in the
([$\alpha$/Fe],\,[Fe/H]) plane in each of 18 bins in the $(R,z)$ plane. The
robustness of this result was recently confirmed by \citet{Eilers2022}. The
upper block of panels in Fig.~\ref{fig:Hayden} is an analogous plot using the
same data, kindly supplied by M.~Hayden. Two
stellar populations are clearly visible. There is a high-$\alpha$ population
that always occupies the same location in the chemical plane but is much more
prominent at large $|z|$ and small $R$: it has essentially disappeared at
$R>10\kpc$ and is very faint at $|z|<3\kpc$. The other, normal-$\alpha$,
population dominates near the plane and at large $R$. It moves to lower
[Fe/H] with increasing $R$.

If we conjecture how the stellar populations of our dynamical model are
distributed in the chemical plane, we can create a analogous figure. The
lower block of panels in Fig.~\ref{fig:Hayden} shows the result of doing this
as follows.

\begin{table*}
\caption{Parameters of the chemical pdfs used to construct
Fig.~\ref{fig:Hayden}. The units of $x,y,\sigma_x,\sigma_y$ are dex, those of
$V_{\rm c}\d\overline{x}/\d J_\phi$ are dex per kpc, those of $\theta$ are
degrees and those of $J_{z0}$ and $\Delta_z$ are $\kpc\kms$.}\label{tab:chem}
\begin{tabular}{lcccccccc}
Component&$\overline{x}_\odot$&$V_{\rm c}\d\overline{x}/\d
J_\phi$&$\overline{y}$&$\sigma_x$&$\sigma_y$&$\theta$&$J_{z0}$&$\Delta_z$\\
\hline
Young disc  &$0.2$&$-0.07$&$   0$&$0.1$&$0.03$&$2$&-&-\\
Middle disc &$  0$&$-0.07$&$0.03$&$0.1$&$0.03$&$2$&-&-\\
Old disc    &$-0.2$&$-0.07$&$0.06$&$0.1$&$0.05$&$2$&-&-\\
Thick disc  &$-0.5$&    0& $0.2$  &$0.2$&$0.03$&$7$&$50$&$15$\\
Stellar halo&$-1$&      0& $0.2$  &$0.3$&$0.1$&$3$&-&-\\
Bulge       &$-0.5$&    0& $0.2$  &$0.2$&$0.02$&$3$&-&- \\
\end{tabular}
\end{table*}

At the centre of each spatial bin we determined the fractional contribution
to the stellar density by stars belonging to each of the six populations,
young disc, middle disc, old disc, thick disc, stellar halo and
bulge. Then mock stars were drawn from the velocity distributions of the
components in the corresponding proportions, the total number of stars being
the same as the number of real stars reported by \cite{Hayden2015} for that
spatial bin. The mock stars were assigned values of [Fe/H] and [$\alpha$,Fe] by
sampling its component's chemical  pdf. Each pdf was a
two-dimensional Gaussian with principal axes
\[\label{eq:chemo_first}
\begin{pmatrix}x\cr y\end{pmatrix}
=\begin{pmatrix}\cos\theta&-\sin\theta\cr\sin\theta&\cos\theta\end{pmatrix}
\begin{pmatrix}\hbox{[Fe/H]}\cr[\alpha/\hbox{Fe}]\end{pmatrix}
\]
 that are rotated with respect to the [Fe/H] and [$\alpha$/Fe] axes.
The angle $\theta$ being small, $x$ is close to [Fe/H] while $y$ is close to
[$\alpha$/Fe].

In the normal-$\alpha$ components, the distribution in $y$ was independent of
$J_\phi$ but the mean value of $x$ was taken to be
\[
\overline{x}=\overline{x}_\odot+{\d\overline{x}\over\d
J_\phi}\left(J_\phi- J_{\phi\odot}\right),
\]
where $J_{\phi\odot}$ is the circular angular momentum at the Sun and
$\overline{x}_\odot$ is approximately the mean metallicity of the component
at $R_0$.  This dependence of $\overline{x}$ on $J_\phi$ generates a
metalliticy gradient $\d\hbox{[Fe/H]}/\d R\simeq V_c\d\overline{x}/\d J_\phi$
in the disc.  

\begin{figure}
\centerline{\includegraphics[width=.9\hsize]{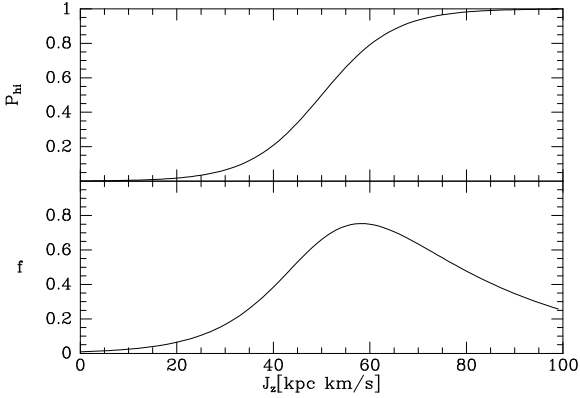}}
\caption{Upper panel: the probability that a thick-disc star has high-$\alpha$
chemistry. Lower panel: the $J_z$-dependence of the DF of the high-$\alpha$
population.}\label{fig:tanh}
\end{figure}

The top row of Fig.~\ref{fig:Hayden} shows that the high-$\alpha$ population
is scarcely visible at $R\ga10\kpc$. The obvious way to engineer this fading
of the high-$\alpha$ is to make the probability $P_{\rm hi}$ of a thick-disc
star belonging to this
population a function of $J_\phi$. However experiments with $P_{\rm hi}$ of
the form
\[
P_{\rm hi}=\fracj12\left[1-\tanh\left({J_\phi-J_{\phi0}
\over\Delta_{\phi}}\right)\right],
\]
produced results that were unsatisfactory in two respects: i) an implausibly
rapid transition was required, specifically $J_{\phi0}/\Delta_\phi\ge100$, and
ii) it left the high-$\alpha$ population too prominent at $z\simeq0.7$ and
$0.2\kpc$. The more satisfactory results shown in the lower block of panels
of Fig.~\ref{fig:Hayden} were
obtained by taking the probability of belonging to the high-$\alpha$
population to be
\[\label{eq:chemo_last}
P_{\rm hi}(J_\phi)=\fracj12\left[1+\tanh\left({J_z-J_{z0}
\over\Delta_{z}}\right)\right].
\]
Now $P_{\rm hi}\simeq1$ for $J_z\gg J_{z0}$ and $P_{\rm hi}\simeq0$ for
$J_z\ll J_{z0}$.  If a star was assigned to the high-$\alpha$ portion of the
thick disc, its chemistry was drawn from a Gaussian in $(x,y)$ independently
of $J_\phi$, while if it was assigned to the normal-$\alpha$ disc, its
chemistry was determined as if it were a member of the old disc. The upper
panel of Fig.~\ref{fig:tanh} is a plot of $P_{\rm hi}(J_z)$ for the chosen
values of $J_{z0}$ and $\Delta_z$.

Table~\ref{tab:chem} lists for each component the numbers that define the
two-dimensional Gaussians pdfs, and the parameters $J_{z0}$ and
$\Delta_{J_z}$ that determine the distribution of high-$\alpha$ stars within
the thick disc. Comparison of the upper and lower blocks of panels in
Fig.~\ref{fig:Hayden} shows that the model captures the essential features of
the observations, including the dominance of the high-$\alpha$ population at
small $R$ and large $z$, and the drift of the normal-$\alpha$ population
towards lower [Fe/H] with increasing $R$.

\section{Discussion}

\subsection{Extracting information from data}

Much previous galaxy modelling has been based on the Jeans equations
\citep{BDI,Roea03,Gaea12,Read2017,Nitschai2020,Nitschai2021,Sivertsson2022}.
Gaia produces such lage samples that the full distribution of velocities in a
small spatial region can be determined with exquisite precision. These
distributions are highly non-Gaussian, so a technique which characterises
them by just two or three moments is throwing away valuable information.

Extraction of the maximum information from a stellar survey is probably
achieved by evaluating the likelihood of the survey data given candidate
models. Unfortunately, this technique is unfeasible when there is significant
obscuration by dust and the density of dust is not well known
\citep[e.g.][]{LiBinneyYoungD}.  Moreover, evaluating likelihoods for
individual stars is very costly for samples as large as those yielded by
Gaia. In light of these considerations there is a compelling case for binning
stars spatially and evaluating the fit to models of the resulting velocity
distributions.  The comparison would be better made by evaluating likelihoods
in velocity space than by binning stars by their single velocity components,
but the cost of likelihood evaluation seems currently prohibitive.
Humans can more easily assess fit quality from one-dimensional velocity
distributions than from three-dimensional ones, but for an MCMC search of
model space one should probably bin stars into cell in velocity space that
are not the two-dimensional slices right across the space employed here.

\subsection{The dark-matter baryon balance}

Hand-fitting models to the data reveals a strong connection between the
radial structure of the dark halo and the extent to which the disc is
self-gravitating. On the one hand, the large distances from the
plane reached by thick-disc stars places an upper limit on the disc's mass.
On the other hand, the clear outwards decline in the circular speed at
$R\ga6\kpc$ requires both a significant contribution to the gravitational
field from stars, and that the circular speed contributed by the dark halo
does not rise strongly in this radial range. The latter requirement places an
upper limit on the core parameter $J_c$, and a lower limit on the index
$\alpha$ of the dark halo's DF. It would be very interesting to use the MCMC
algorithm to explore how big a region of model space acceptable models
occupy.

\subsection{We need better halo DFs}

Action-based DFs that envisage a non-zero density of stars in the
neighbourhood of $J_\phi=0$ need to conform to certain constraints
\citep{PifflPenoyreB,Binney_in_prep}. The disc
DFs introduced here have vanishing star densities at $J_\phi=0$ so do not need
to engage with these constraints, but the spheroidal DFs of the dark halo, the bulge and
the stellar halo should conform to these constraints, and they do so to only
a limited extent. To avoid excessive violation of the constraints, and
the introduction of unphysical features, the parameters of the spheroidal DFs
have been chosen to avoid plausible levels of radial bias. The absence of
radial bias in the dark halo and the bulge is not significant observationally
because the data contain few bulge stars and no dark-matter particles. But
the stellar halo needs to have greater radial bias and work to enable this
should be a high priority. Moreover, restricting the velocity anisotropy of
the dark halo's DF limits the dark halo's flattening, and a flattened halo
will imply a less massive disc. Hence, better functional forms for spheroidal
DFs may significantly change our understanding of the disc-halo balance.

\subsection{Structure of the high-$\alpha$ population}

The quality of the fit to the \cite{Hayden2015} data that is provided by the
restriction of high-$\alpha$ stars to orbits with large $J_z$ is so good that
it seems certain that the restriction is real, but its historical origin is
far from evident. The {\it a priori} expectation is that high-$\alpha$ stars would
be restricted to low $|J_\phi|$ because these would be the first stars to form
in the natural inside-out formation picture. However, fruitless attempts to
model the data in this way left us convinced that it is to orbits with large
$J_z$ that high-$\alpha$ stars are confined. 

In light of this result the high-$\alpha$ population would form a very
strange component if considered in isolation, for it comprises stars that
have apocentres in $z$ that fall in quite a narrow range -- the lower panel
of Fig.~\ref{fig:tanh} illustrates this fact by showing how the DF of this
populaton depends on $J_z$. It seems that at low $|z|$ the density of high-$\alpha$ stars
{\it increases} with $z$, peaking where a star with $J_z\simeq50\kpc\kms$
turns around.

Could the $J_z$-distribution of the high-$\alpha$ stars have arisen when a
primordial, high-$\alpha$, thin disc was shattered in a merger
\citep{Belokurov_splash}? The problem
with this idea is that we would expect the merger to spread stars in
$J_z$ rather than shift them all up in $J_z$, leaving orbits around $J_z=0$
vacant until occupied by the subsequent formation of low-$\alpha$ stars. A
possibly more promising scenario that the high-$\alpha$ stars were stripped from a
satellite that was finally disrupted  on a nearly circular orbit, and their characteristic
value of $J_z\sim50\kpc\kms$ was the vertical action of the satellite when it
dispersed. Another possibility is that we are seeing the result of
`levitation' \citep{SridharTouma1996}: stars trapped in a resonance such as
$\Omega_r=\Omega_z$ moving up in $J_z$ along with the resonance as the disc
gains mass and the values of $\Omega_z/\Omega_r$ in the plane increase.

\subsection{What to do with all Gaia stars}

Gaia is a survey
instrument that provides photometry of unprecedented precision. Hence it is
the ultimate tool with which to determine how the density of stars varies
with location. Yet we have completely neglected this possibility and instead
used a forty-year-old determination of the density of stars above the Sun
\citep{GiRe83}. The problem is that the Gaia catalogues have complex
selection functions. Even after these selection functions have been
determined \citep{EverallBoubert2022}, to predict how many stars will be
catalogued at a given sky position one needs accurate knowledge of the
extinction as a function of distance along the relevant line of sight. This
knowledge is currently woefully lacking for most sight lines, but fortunately
along lines of sight at $|b|\ga80\,$deg extinction is sufficiently low that a
poor understanding of its distribution is not a major problem, so it {\it is}
possible to deduce the vertical profile $\rho_*(z)$ of the disc near the Sun
from Gaia data, and \cite{Everall_za,Everall_zb} have recently done this.
Their work appeared after the present work was largely completed and a
decision was made not to replace the \cite{GiRe83} data with the new
determination of $\rho_*(z)$ because it would be fundamentally sounder to fit
the self-consistent models directly to Gaia's star counts rather than to a
density profile that has been fitted to the star counts --
\cite{Everall_za,Everall_zb} needed to assume that the luminosity function is
independent of $z$, though it will not be because younger populations are
more prominent at low $|z|$, and older and more metal-poor populations
dominate at high $|z|$. 

Since each population of the current models
comes with an age range and therefore a luminosity function, predicting the
star counts along a dust-free line of sight is straightforward. Comparison of
these star counts, modified by the relevant Gaia selection function, with the
actual star counts will then yield a constraint on the models that can
replace that provided here by Fig.~\ref{fig:rhoz}.

Probably the best way to use comparisons between predicted and actual Gaia
star counts along general lines of sight is for the construction of a global
three-dimensional dust map. Current attempts to construct dust maps are
restricted to the tiny fraction of stars for which it is possible to obtain
an extinction \citep{Sale2014,Green2018,Green2019,Lallement2022}. Such stars, mainly intrinsically
blue stars, are for some reason susceptible to having their absolute
magnitudes predicted. A model of the type presented here, combined with a
trial dust model, yields for each field of view  star counts as a function
of parallax that can be compared with the actual star counts down to the
faintest magnitudes. Hence every star in the Gaia catalogue can be used to
constrain the dust model.

\section{Conclusions}\label{sec:conclude}

Fully self-consistent dynamical galaxy models constitute valuable tools for the
interpretation of observational data for both our Galaxy and external
galaxies because they pull constraints from a variety of
observational probes into a coherent physical framework. They minimise the
number of parameters that must be determined from observations by exploiting
to the full the constraints imposed by the laws of dynamics.

Such models are inherently steady-state models, and currently \agama\ can
only construct axisymmetric models. However, from such a model one may easily
draw an N-body sample, and by perturbing this either study time-dependent
phenomena \citep[e.g.][]{BinneySchoenrich2018,AlKazwini2022} or construct
barred models by the made-to-measure technique
\citep{SyerTremaine,deLorenzi2007}.

Hitherto, self-consistent models of our Galaxy
\citep{PifflPenoyreB,BinneyPiffl15,ColeBinney} have represented the discs
with the quasi-exponential DF. Unfortunately, the radial and vertical
epicycle frequencies $\kappa(J_\phi)$ and $\nu(J_\phi)$, and the circular
radius $R_{\rm c}(J_\phi)$ play significant roles in this DF, so the DF is
only fully specified when a potential is given. This fact makes the
quasi-isothermal DF ill-suited to self-consistent galaxy modelling because
the galaxy's potential should emerge from the DF, not precede it.

Therefore in Section~\ref{sec:discDF}
we introduced a new family of DFs for discs that are fully
specified by the disc's mass and the values of seven parameters. We have
elucidated the physical significance of the parameters: three characteristic
actions set the scale length of the disc and the in-plane and vertical
velocity dispersions.  Two further parameters set the radial gradients of the
dispersions and the final two control the central structure of the disc.

In Section~\ref{sec:models} we  built a model of our Galaxy that
comprises four of these stellar discs, a gas disc, and three spheroidal
components. The model reproduces to good accuracy the velocity distributions
of stars in the Gaia RVS sample at 35 locations distributed through the
rectangle $R_0\pm3\kpc$ and $|z|\le3\kpc$. It also fits old observations of
the density of stars as a function of $|z|$ in the column above the Sun,
which effectively set the balance between baryon and dark mass. 

The model's circular-speed curve is tightly constrained over the $\sim6\kpc$
radial range covered by the Gaia RVS sample and it agrees well with recent
determinations using specific tracer populations.  Its stellar disc has mass
$3.9\times10^{10}\msun$, scale length $2.6\kpc$ and local surface density
$22\msun\pc^{-2}$. Its dark halo has mass $M_{200}=80.5\times10^{10}\msun$,
scale radius $r_s=20\kpc$ and local density $0.012\msun\pc^{-3}$. Its original
central cusp is assumed to have been eroded by interactions with baryons. We
reconstructed the  dark halo prior to baryon infall. Prior to
the addition of baryons it had peak circular speed $170\kms$ and virial
radius
$r_{200}=192\kpc$.

The model's stellar disc is a superposition of three relatively cool ``thin''
discs of increasing velocity dispersions and therefore thickness, and a much
hotter and thicker disc. The disc's scale lengths decrease as their
dispersions and thicknesses increase. The hotter a disc is, the faster its
velocity dispersions fall off with increasing radius.  

After assigning a
chemical composition to each of the model's six stellar components, we could
predict from the model how the chemistry of stars varies with location in the
$Rz$ plane. We identified simple compositions of the components that yielded
good agreement between the model's predictions and the chemistry measured by
the APOGEE survey \citep{Hayden2015}. The key feature of these compositions
is the restriction of high-$\alpha$ stars to orbits with high $J_z$. This
restriction ensures that normal-$\alpha$ stars dominate both at low $|z|$ and
small $R$, and at high $|z|$ and large $R$.

The Galaxy model presented here is designed mainly to illustrate the
opportunities that are opened up by the new disc DFs. There is much scope to
extend the work by a wider exploration of model space and the addition of
more and better observational constraints.

Here we have presented a single model that provides good fits to most of the
data. Attempts to find a best-fitting model were unsuccessful on account of
the high dimensionality of the model parameter space and the cost of
computing observables from a model. Nevertheless, soon the MCMC
algorithm should be used to explore model space starting from this model and
thus to obtain insight into the range of model parameters that generate
acceptable models.

\section*{Acknowledgements}

This work was supported by the UK Science and Technology Facilities Council
under grant number ST/N000919/1.  JB also acknowledges support from the
Leverhulme Trust through an Emeritus Fellowship. We thank Michael Hayden for
providing the APOGEE data re-plotted in Fig.~\ref{fig:Hayden}.

This work presents results from the European Space Agency (ESA) space mission
Gaia. Gaia data are being processed by the Gaia Data Processing and Analysis
Consortium (DPAC). Funding for the DPAC is provided by national institutions,
in particular the institutions participating in the Gaia MultiLateral
Agreement (MLA). The Gaia mission website is https://www.cosmos.esa.int/gaia.
The Gaia archive website is https://archives.esac.esa.int/gaia.

\section*{DATA AVAILABILITY}
The code that generates Galaxy  models can be downloaded from the \agama\
website https://github.com/GalacticDynamics-Oxford/Agama

\def\physrep{Phys.~Reps}
\bibliographystyle{mn2e} \bibliography{/u/tex/papers/mcmillan/torus/new_refs}

\end{document}